\documentclass[manuscript,a4paper,article]{aastex}
\usepackage{bibentry}
\usepackage{cite}
\usepackage{amsmath}
\usepackage{color}
\usepackage{ulem}
\usepackage{graphicx}
\renewcommand{\raggedright}{\leftskip=0pt \rightskip=0pt plus 0cm}
\raggedright
\newcommand{\Chandra}{{\it Chandra}}

\newcommand{\XMM}{{\it XMM-Newton}}
\newcommand{\ROSAT}{{\it ROSAT}}
\newcommand{\ASCA}{{\it ASCA}}
\newcommand{\Einstein}{{\it Einstein}}

\usepackage{epsfig} 
\graphicspath{{figures/}}
\begin{document}

\title{A \Chandra\ Study of the Image Power Spectra of 41 Cool Core and Non-Cool Core Galaxy Clusters}

\author{Chenhao Zhang\altaffilmark{1}, Haiguang Xu\altaffilmark{1,2}, Zhenghao Zhu\altaffilmark{1},  Weitian Li\altaffilmark{1}, Dan Hu\altaffilmark{1}, Jingying Wang\altaffilmark{4}, Junhua Gu\altaffilmark{4}, Liyi Gu\altaffilmark{5}, Zhongli Zhang\altaffilmark{6}, Chengze Liu\altaffilmark{1}, Jie Zhu\altaffilmark{3} \scriptsize AND \normalsize Xiang-Ping Wu\altaffilmark{4}}

\altaffiltext{1}{Department of Physics and Astronomy, Shanghai Jiao Tong University, 800 Dongchuan Road, Minhang, Shanghai 200240, China; email: eotss813@sjtu.edu.cn, hgxu@sjtu.edu.cn;}

\altaffiltext{2}{IFSA Collaborative Innovation Center, Shanghai Jiao Tong University, 800 Dongchuan Road, Minhang, Shanghai 200240, China}

\altaffiltext{3}{Department of Electronic Engineering, Shanghai Jiao Tong University, 800 Dongchuan Road, Minhang, Shanghai 200240, China.}

\altaffiltext{4}{National Astronomical Observatories, Chinese Academy of Sciences, 20A Datun Road, Beijing 100012, China;}

\altaffiltext{5}{SRON Netherlands Institute for Space Research, Sorbonnelaan 2, 3584 CA Utrecht, the Netherlands}

\altaffiltext{6}{Max-Planck-Institut f{\"u}r Astrophysik, Karl-Schwarzschild-Stra$\ss$e 1, 85740 Garching, Germany}

\begin{abstract}
In this work we propose a new diagnostic to segregate cool core (CC) clusters from non-cool core (NCC) clusters by studying the two-dimensional power spectra of the X-ray images observed with the \Chandra\ X-ray observatory. Our sample contains 41 members ($z=0.01\sim 0.54$), which are selected from the \Chandra\ archive when a high photon count, an adequate angular resolution, a relatively complete detector coverage, and coincident CC-NCC classifications derived with three traditional diagnostics are simultaneously guaranteed. We find that in the log-log space the derived image power spectra can be well represented by a constant model component at large wavenumbers, while at small wavenumbers a power excess beyond the constant component appears in all clusters, with a clear tendency that the excess is stronger in CC clusters. By introducing a new CC diagnostic parameter, i.e., the power excess index (PEI), we classify the clusters in our sample and compare the results with those obtained with three traditional CC diagnostics. We find that the results agree with each other very well. By calculating the PEI values of the simulated clusters, we find that the new diagnostic works well at redshifts up to 0.5 for intermediately sized and massive clusters with a typical \Chandra\ or \XMM\ pointing observation. The new CC diagnostic has several advantages over its counterparts, e.g., it is free of the effects of the commonly seen centroid shift of the X-ray halo caused by merger event, and the corresponding calculation is straightforward, almost irrelevant to the complicated spectral analysis. 
\end{abstract}


\section{INTRODUCTION}
X-ray observations performed in the past two decades have revealed that more than half of the galaxy clusters host a bright, dense core where the intracluster medium (ICM) has cooled down to temperatures lower than that of the ambient gas, so that such cool core (CC; a nomenclature proposed by \citealt{MP01}) clusters usually exhibit sharply peaked central X-ray emission. For example, based on the deprojected imaging analysis of an \Einstein\ sample of 207 clusters \citet{white97} found that cool cores appear in $62^{+12}_{-15}\%$ of the clusters. Almost at the same time \citet{peres98} studied a \ROSAT\ sample of 55 clusters and estimated that $70-90\%$ of the sample members can be classified as cool core systems. Later \citet{chen07} presented an imaging spectroscopic study of both \ASCA\ and \ROSAT\ data of 106 clusters, which were drawn from the HIFLUGCS (the Highest X-ray Flux Galaxy Cluster Sample; \citealt{reiprich01}; \citealt{RB02}) sample, and concluded that about 49\% of the sample clusters have cool cores. Recently by analyzing the \Chandra\ data of a statistically complete sample of 64 X-ray selected HIFLUGCS clusters \citet{hudson10} found that the chance of the presence of cool core is 72\% (the different claimed CC-fractions is mainly as a result of different definitions of CC, e.g., see \citet{hudson10} for study of different CC-diagnostics). Despite the fact that all these samples whose redshifts are mostly within 0.2, are flux-limited and thus may be biased to a certain degree toward clusters with a bright core, it is very clear that the cool core can be regarded as a common phenomenon in clusters located at $z \lesssim 0.2$.

The characteristic radiative cooling time of a typical cool core is shorter than the age of the cluster, thus a significant cooling flow of a mass deposition rate of $\sim 10^{2-3}$ $\rm M_{\sun} ~ yr^{-1}$ should have been operating as predicted by the traditional cooling flow model \citep[e.g.,][]{fabian94}. However, the observed X-ray gas temperatures of the cool cores never fall below a few keV, and much less cooling signature is found than if cooling-flows progressed unimpeded in radio, infrared, optical, and UV bands for the gases postulated to have a broad spectrum of temperature, especially those cooled rapidly to $< 10^{4}$ K. The contradiction between theory and observation invokes additional heating mechanisms (see \citealt{makishima01}, \citealt{MN07}, and \citealt{fabian12} for a comprehensive review), among which the heating provided by Active Galactic Nuclei (AGN) is the most prevalent one since it is both adequately energetic and self-regulated. Actually a clear CC-AGN connection has been suggested via the detections of X-ray cavities, jets, lobes, and probably radio mini-halos (radio mini-halos are not unambiguously linked to the AGN, e.g., \citealt{feretti12}) surrounding the radio-loud brightest cluster galaxies (BCGs) on roughly the same scales of the cool cores \citep[e.g.,][]{fabian06,wise07,baldi09,blanton11}. In about 70\% of the CC clusters the central dominating galaxy appears as a radio galaxy \citep[e.g.,][]{burns90,DF06,best07,mittal09}, and in nearly all CC clusters a radio emitting AGN creating cavities in the X-ray gas is found \citep[e.g.,][]{burns90,eilek04,sanderson06,fabian12}, notice that the radio-fraction depends on the strength of the CC (i.e. fraction increases from $\sim 67\%$ to $\sim 100\%$ for weak CC clusters to strong CC clusters, \citealt{mittal09}). On the other hand, \citet{sun09} found that in a \Chandra\ sample containing 152 groups and clusters all 69 BCGs with a 1.4 GHz power exceeding $2 \times 10^{23}$ W~Hz$^{-1}$ possess cool cores. Therefore, it is natural to speculate that the AGN outbursts suppress the overcooling of a CC cluster's core region and maintain the cluster in the CC state, meanwhile the feedback from the cooling gas in the cool core fuels and regulates the AGN activity \citep[e.g.,][]{donahue05}.

Although the scenario that AGN and cool core interact with each other in an adaptive way is self-consistent, it does not fully agree with the observations of non-cool core (NCC) clusters, where the cool core has been prevented from forming and the fueling of the AGN is expected to have been quenched. The first counterintuitive case was reported by \citet{gastaldello08}. The authors found that in the relaxed poor cluster AWM 4 the gas temperature shows a relatively flat distribution at $\sim 2.6$ keV and drops outwards quickly at $\gtrsim 200$ kpc,  inferring that the cool core has been erased by a a past, major heating episode. This is supported by the facts that the corresponding central cooling time ($\simeq 3$ Gyr) is long and the central gas entropy is high. Even with the newest high-quality \Chandra\ data only a tiny cool core-like feature or a galactic corona can be identified within the innermost 2 kpc \citep{sun09,osullivan10}. In contrast to A1650 and A2244 that show similar thermodynamic properties \citep{donahue05}, however, AWM 4 harbors an intermediately active central radio galaxy with extended radio lobes out to 100 kpc, which indicates that NCC system can also possess an AGN. In fact similar AGN outbursts also appear in about 45\% of the NCC systems (\citealt{mittal09}; see also \citealt{sun09} and references therein). Apparently these results expose our poor understanding about the CC-AGN relationship, raising the challenging question as to whether there exists a distinct difference between CC and NCC clusters; in other words, is the dichotomal classification of CC and NCC systems intrinsically reliable? To answer the question a deeper comparison between the X-ray properties of CC and NCC clusters is apparently necessary.

In order to define the CC-NCC dichotomy different diagnostics have been proposed in terms of, e.g., central temperature drop \citep[e.g.,][]{sanderson06,burns08}, central cooling time \citep[e.g.,][]{bauer05,ohara06,donahue07}, surface brightness concentration \citep[e.g.,][]{santos08}, mass deposition rate \citep[e.g.,][]{chen07}, or X-ray surface brightness cuspiness \citep[e.g.,][]{vikhlinin07}. To determine which one of these can be used to unambiguously segregate CC from NCC clusters \citet{hudson10} applied 16 CC diagnostics to a sample of 64 HIFLUGCS clusters ($z \lesssim 0.2$), and found that the central cooling time is the best diagnostic parameter for nearby clusters with high quality data, whereas the cuspiness is the best for high redshift ($z \gtrsim 0.03$) clusters. In this work we further address this issue by introducing a new CC diagnostic based on the study of the X-ray image power spectra, which may provide an exquisitely detailed view of the imaging information \citep[e.g.,][]{walker15,zhuravleva15}, of 41 galaxy clusters ($z=0.01\sim 0.54$) selected from the \Chandra's 15-year data archive. The paper is organized as follows. In \S 2 we describe sample selection criteria and data preparation. In \S 3, we present data analysis and calculation of image power spectra. In \S 4 and \S 5 we discuss our results and summarize the work, respectively. Throughout the paper we adopt a flat $\Lambda$CDM cosmology with density parameters $\Omega_{m} = 0.27$ and $\Omega_{\Lambda} = 0.73$, and Hubble constant $\textrm{H}_0 = 71\ \mathrm{km\textbf{ }s}^{-1}\textbf{ }\mathrm{Mpc}^{-1}$. Unless stated otherwise, we adopt the solar abundance standards in \citet{GS98} and quote errors at 68\% confidence level.

\section{SAMPLE SELECTION AND DATA PREPARATION}
In order to characterize the power spectrum of the X-ray image of a galaxy cluster, a high photon count, an adequate angular resolution, and a relatively complete detector coverage of the cluster should be guaranteed simultaneously. Therefore we constructed our sample by searching the \Chandra\ archive for all public pointing observations\footnote{until October 1st, 2015} of galaxy clusters that satisfies the following three criteria: (1) the cluster was observed out to at least $0.45r_{500}$ ($r_{500}$ is defined as the radius within which the mean density of the enclosed gravitating mass is 500 times the critical density of the universe at the cluster's redshift; see \S 3.2.3), meanwhile the $\lesssim 0.35r_{500}$ regions is fully or nearly fully covered by the S3 or I0-3 CCDs of the \Chandra\ Advanced CCD Imaging Spectrometer (ACIS), (2) the number of the photons collected within $< 0.45r_{500}$ during the observation is more than 12500 cts, and (3) the cluster should have been classified explicitly as a strong cool core (SCC), weak cool core (WCC), or non-cool core (NCC) system coincidentally with three traditional CC diagnostics (i.e., diagnostics based on the calculations of central cooling time, cuspiness, and concentration parameter; \S3.2.4). With these selection criteria we selected 41 galaxy clusters as the sample members, whose basic properties are listed in Table \ref{tbl-1}.

For each cluster we started from the ACIS level-1 event files and followed the standard \Chandra\ data processing procedure to reduce the data with CIAO v4.4 and CALDB v4.4.8. To be specific, we first excluded the bad pixels, bad columns, and events with \ASCA\ grades 1, 5 and 7. Next we carried out corrections for the gain, charge transfer inefficiency (for the observations performed after January 30, 2000), astrometry, and cosmic ray afterglow. By examining the light curve extracted in $0.5-12$ keV from source-free regions or regions slightly affected by the sources, using \textbf{LC\_CLEAN} scrip in Sherpa we identified and removed the data contaminated by occasional particle background flares during which the count rate is increased by $20\%$ over the mean value. In addition to above, we masked all the point sources detected beyond the $3\sigma$ threshold in the ACIS images with CIAO tools \textbf{celldetect} and \textbf{wavdetect} (\citealt{freeman02}).

\section{DATA ANALYSIS AND RESULTS}
\subsection{Background}
In order to construct a local background template for each observation, we extract the spectrum from the boundary regions on the S3 CCD or I0-3 CCDs, where the influence of the thermal emission of the intracluster medium (ICM) is relatively weak, and fit the extracted spectrum with a model that consists of the ICM emission (a thermal APEC component absorbed by the Galactic column density given in \citealt{kalberla05} and \citealt{DL90}. The gas abundance is fixed to $0.3$ $Z_\odot$ if it is not well constrained), the Cosmic X-ray Background (CXB; a power-law component with $\Gamma=1.4$, which is also absorbed by the Galactic column density; e.g., \citealt{mushotzky00}; \citealt{CR07}), the Galactic emission (two APEC components with $kT = 0.2$ keV and $0.08$ keV, respectively; e.g., \citealt{HB06}; \citealt{gu12}), and the particle-induced hard component derived from the corresponding \Chandra\ blanksky templates available at the \Chandra\ Science Center. The background template for the observation (i.e., Galactic + CXB + particle components) can thus be determined when the best-fit is achieved. When the corresponding data are available, we have also compared the count rate of our background template calculated in $0.2-2$ keV, where the effect of the particle component is less significant, with that of the archive \ROSAT\ All-Sky Survey (RASS) diffuse background maps, and obtained consistent results. It is difficult to estimate the field-to-field variations of both the Galactic and CXB background components in each observation. As an approximation in the analysis that follows we estimate the model parameter error ranges by taking into account both statistical and systematic uncertainties (10\% as a conservative estimate; \citealt{kushino02}) in the background.

\subsection{Imaging Spectroscopic Study of Gas Properties }
\subsubsection{Gas Temperature Distributions}
For each observation we define $5-7$ concentric annuli, which are all centered on the cluster's X-ray peak. The width of each annulus is determined to guarantee a minimal photon count of 2500 cts in $0.7-7$ keV, and for the outermost annulus the condition that the photon count is at least twice the background should be satisfied simultaneously. We extract the \Chandra\ ACIS S3 or I0-3 spectra from these concentric annuli, and fit them by using the X-ray spectral fitting package XSPEC v12.8.2 (\citealt{arnaud96}). To minimize the effects of the instrumental background at higher energies and the calibration uncertainties at lower energies, we limit the fittings in $0.7-7$ keV. In the model fits, we estimate the influence of the outer spherical shells on the inner ones by using the XSPEC model PROJECT, and fit the deprojected spectra with the optically thin, collisional plasma model APEC (\citealt{smith01}), which is absorbed by the foreground photoelectric absorption model WABS (the column density $N_{\rm H}$ is fixed to the corresponding Galactic value; see \citealt{kalberla05} and \citealt{DL90}). Whenever the gas metal abundance is not well constrained, we fix it to 0.3 $Z_\odot$. We add an additional absorbed APEC component for the innermost annulus since it might be contaminated by the coexisting multi-phase gases (e.g., \citealt{makishima01}), if the F-test shows that the fitting is improved at the 90\% confidence level. In this case the ICM temperature is defined as that of the hot phase. Smoothed gas temperature profiles, as well as the smoothed metal abundance profiles, are then derived by running cubic spline interpolation to the best-fit model parameters for the annulus set. Finally we calculate the average gas temperature of the cluster (Table~\ref{tbl-1}) by fitting the spectra extracted between $ 0.2-0.5 r_{500}$ (\S 3.2.3) using the same spectral model as above.

\subsubsection{X-Ray Surface Brightness Profiles and Gas Density Distributions}
We create the exposure map for each observed image by using the spectral weights calculated for an incident thermal gas spectrum that possesses the same average temperature and metal abundance as the cluster (\S 3.2.1). After the X-ray images are corrected by applying the exposure maps to remove the effects of vignetting and exposure time fluctuations, we use concentric annular bins, which are all centered at the X-ray peak of the gas halo, to extract the X-ray surface brightness profiles $S_X(R)$ ($R$ is the two dimensional radius) in $0.7-7$ keV. 

Under the assumptions of hydrodynamic equilibrium and spherical symmetry, the three-dimensional distribution of gas electron density $n_e$ can be expressed with a $\beta$-model \citep[e.g.,][]{cavaliere76}, 
 	\begin{equation}
 	n_e(r)=n_0\left[1+\left(\frac{r}{r_c}\right)^2\right]^{-3\beta/2},
 	\end{equation}  
or a double-$\beta$ model \citep[e.g.,][]{jones84}
 	\begin{equation}
 	n_e(r)=n_{0,1}\left[1+\left(\frac{r}{r_{c,1}}\right)^2\right]^{-3\beta_1/2}+n_{0,2}\left[1+\left(\frac{r}{r_{c,2}}\right)^2\right]^{-3\beta_2/2}
 	\end{equation} 
when a detectable central surface brightness excess appears in the inner regions, where $r_c$ is the core radius and $\beta$ is the slope. With the derived gas density profile and the profiles of gas temperature $T(r)$ and metal abundance $A(r)$, which are obtained in \S3.2.1, we model the X-ray surface brightness profile as 
 	\begin{equation}\label{s_fit}
 	S_X(R)=\int_{R}^{\infty}\Lambda(T,A)n_en_p(r)\frac{rdr}{\sqrt{r^2-R^2}}+S_{\rm bkg},
 	\end{equation}
where $S_{\rm bkg}$ is the background, and $\Lambda(T,A)$ is the cooling function. The density $n_e(r)$ is determined when the best-fit to the observed surface brightness profile is achieved via $\chi^{2}$-test.

\subsubsection{Characteristic Radius $r_{500}$}
In order to derive the characteristic radius $r_{500}$, we first calculate the distribution of the total gravitating mass of the cluster in the regions covered by \Chandra{}'s field of view under the hydro-statics equilibrium assumption
 	 \begin{equation}
 	 M(<r)=-\frac{r^2k_bT_X}{G\mu m_p}\left[\frac{1}{T_X}\frac{dT_X}{dr}+\frac{1}{n_e}\frac{dn_e}{dr}\right],
 	 \end{equation}
where $\mu$ $=$ $0.61$ is the mean molecular weight per hydrogen atom, $k_b$ is the Boltzmann constant, and $m_p$ is the proton mass. In the regions outside \Chandra{}'s field of view, the mass profile $M(<r)$ is obtained by fitting the result derived above by applying the NFW profile \citep{navarro96}   
 	 \begin{equation}
 	  \rho(r)=\frac{\rho_{0}}{\left(1+r/r_s\right)^2r/r_s},
 	 \end{equation}
where $\rho(r)$ is the density of the total gravitating mass, and extrapolating the best-fit mass profile out to the point where the mean density of the enclosed gravitating mass is $500$ times the critical density of the universe at the cluster's redshift (Calculated $M_{500}$ and $r_{500}$ are listed in Table~\ref{tbl-1}). The comparison of $M_{500}$ between this and previous work (\citealt{zhao15}) was shown in Figure 1, the results are consistent with each other.

\subsubsection{SCC-WCC-NCC Classifications with Traditional Diagnostics}
Using the observed profiles of X-ray surface brightness, gas temperature, metal abundance and gas density, the clusters in our sample can be classified explicitly as an SCC, WCC, or NCC system coincidentally\footnote{In other words, a cluster is included in the sample only when the three traditional diagnostics give the same SCC/WCC/NCC classification} with three frequently quoted traditional CC diagnostics, i.e., the central cooling time (CCT, Hudson et al. 2010)
\begin{equation}
t_{\rm cool} \equiv \frac{5}{2} \frac{(n_e+n_i) kT}{n_e n_H \Lambda(T,A)} \quad {\rm for} \quad r<0.048r_{500},
\end{equation}
cuspiness (Vikhlinin et al. 2007)
\begin{equation}
\alpha \equiv - \frac{d \rm log(n_e)}{d \rm log(r)} \quad {\rm at} \quad r=0.04r_{500},
\end{equation}
and surface brightness concentration parameter (Santos et al. 2008)
\begin{equation}
C_{\rm SB} \equiv \frac{\sum(r\leq40 {\rm kpc})}{\sum(r\leq400 {\rm kpc})},
\end{equation}
which is the ratio of the integrated surface brightnesses within central 40 kpc to that within 400 kpc. With these diagnostic parameters and the corresponding criteria listed in Table~\ref{tbl-2}, the sample clusters are classified as an SCC, WCC, or NCC system, as shown in column 8 before the slash mark in Table~\ref{tbl-3}.

We find that in nine out of 12 SCC clusters the ratio of the central luminosity excess to the total luminosity integrated within $0.45r_{500}$ ($R_{\rm excess} \equiv L_{\rm excess}^{\rm 0.7-7 keV} / L_{\rm total}^{\rm 0.7-7 keV}$) is significant, ranging from about $8\%$ to $50\%$ (Table~\ref{tbl-3}). In seven out of nine WCC clusters a central surface brightness excess is detected, but the ratio $R_{\rm excess}$ is much lower (typically a few percent) except in A1651 ($\simeq 10\%$) and ESO306-G170B ($\simeq 21\%$). In NCC clusters the effect of the central luminosity excess, if it does exist, is actually negligible.

\subsubsection{Power Spectra of the X-Ray Images}
For each cluster we first calculate the two-dimensional Fourier transform of the background-subtracted and exposure-corrected X-ray image (i.e., the flux distribution $F({\vec R})$), which is not smoothed, as
\begin{equation}
\tilde{F}(\vec{k})=\int_{S} F(\vec{R})e^{i\vec{R}k}d\vec{F},
\end{equation}
where $k$ is the wavenumber, and $S$ represents the hole image by using the MATLAB tool fft2 and fftshift. We removed all the point sources detected beyond $3\sigma$ threshold, and fill the corresponding regions via interpolation with neighboring pixels assuming Poisson statistics. The CCD gaps between the ACIS-I chips are filled in the same way. We find that the systematic errors introduced in the process are much smaller than the statistical errors, which will be estimated below, at wavenumbers that we are interested in ($0.2$ ${\rm kpc}^{-1} \leq k \leq 0.001$ ${\rm kpc}^{-1}$ ).    
 The two-dimensional power spectrum is then obtained as
\begin{equation}
P(k)=<|\tilde{F}(\vec{k})|^2>, 
\end{equation}
Next, we create a random fluctuation distribution $F_{\rm err}({\vec R})$ for the observed image by running Monte-Carlo simulations to estimate the fluctuation on each pixel, which is assumed to follow the Poisson distribution, and add the fluctuations map into the observed map to create a simulated image $F_{\rm sim}({\vec R})$ (i.e., $F({\vec R}) + F_{\rm err}({\vec R})$). After 100 simulated images are randomly created, we calculate their power spectra and use the scatter of them to determine the error range of the image power spectrum $P(k)$ of the cluster. In the simulation we have added an additional 5\% systematic error to account for the subtle numerical and instrumental effects (e.g., errors introduced by the griding of image, the periodic boundary conditions, the inhomogeneity in the exposure between neighboring pixels, and the method of removing and filling point sources; \citealt{hudson10}). The obtained power spectra and the corresponding error ranges for the sample clusters are plotted in Figures~\ref{fig2}.

We find that in the log-log space the derived image power spectra of $39$ out of $41$ sample clusters become flat as the wavenumbers increases and can be represented by a constant component when the wavenumber is large enough ($ k \geq 0.01-0.05$ $\rm kpc^{-1}$ for $31$ clusters, and $ k \geq 0.05-0.1$ $\rm kpc^{-1}$ for seven clusters, $ k \geq 0.1$ $\rm kpc^{-1}$ for one cluster), while at small wavenumbers a power excess beyond the constant component appears in all clusters, with a clear tendency that the power excess is stronger in CC clusters than in NCC clusters. This is very likely due to the fact that CC clusters usually harbor large bright cores \citep{vikhlinin07,santos08}, which may significantly enhance the power on scales comparable to the cooling radius ($r_{\rm cool}$, $\sim 0.1r_{500}$ as shown in Table~\ref{tbl-3}; cooling radius is the radius within which the gas cooling time is $< 7.7\times10^{9}$ yr, the light travel time form $z=1$; \citealt{rafferty06}). To illustrate this we plot the relative core brightness (defined as the ratio of integrated surface brightness within central 0.048$r_{500}$ to that within 0.45$r_{500}$) against the power ratio (defined as the ratio of power at 0.35$r_{500}$ to that at large wavenumbers where image power spectra reduce to a constant component) which reflect the power excess we mentioned above in Figure~\ref{fig3}, and find that there indeed exists an unambiguous correlation between the two parameters. We also have attempted to fit the observed power spectra by adding a $\beta$ component into the model to represent the power excess, and find that in all sample clusters the model ($\beta$+constant component) can provide an acceptable fitting to the power spectrum in the log-log space (Fig.~\ref{fig2}). On the other hand, the model that consists of two $\beta$ components always gives a relatively poor fit to the observed power spectrum in nearby all galaxy clusters.

In Figure~\ref{fig4} we plot the two-dimensional power spectra of the X-ray images of all sample clusters, together with the same power spectrum set scaled by $r_{500}$ (x-axis) and the power at $0.01r_{500}$ (y-axis). In all cases the power spectra is truncated at five $\rm pixel^{-1}$. In the figures we color-code the power spectra according to the classification based on the traditional cool core diagnostic parameters (i.e., central cooling time, cuspiness, and concentration parameter; see \S 3.2.4 and Table~\ref{tbl-2}): blue for SCC clusters, green for WCC clusters, and red for NCC clusters. The errors are typically about 10\% for small wavenumbers and 20\% for large wavenumbers, which consistent with the errors of \citet{zhuravleva15} in the same scale (\citet{zhuravleva15} specifically studied Perseus cluster), and are not shown in the figures for a better visualization of the profiles. Despite the large scatter, we find that there are systematic distinctions between the shapes of the unscaled power spectra of SCC, WCC, and NCC systems. The distinctions become more obvious when we examine the scaled power spectra, which implies that the \Chandra{} image power spectra may be used to classify galaxy clusters located with in z $\leq 0.5$ as CC or NCC systems, as will be discussed below.

\section{DISCUSSION}
In order to characterize the distinctions between the image power spectra of SCC, WCC, and NCC systems, for each cluster we locate two specific points on the line of the scaled power spectrum at wavenumbers $k=1/0.35$ and $1/0.035$, which correspond to the scales of $0.35r_{500}$ and $0.035r_{500}$ in the real image space, respectively, and use the two points as the left-top and right-bottom vertices to determine a rectangular box (Fig.~\ref{fig5})\footnote{The positions of the points are chosen to cover the range where the observed power spectrum turns to be flat as the wavenumber increases}. Using this box we introduce a new CC diagnostic parameter, i.e., the power excess index (PEI), to quantify the degree of the power excess as 
$\Delta_{\rm PE} \equiv A_{\rm below} / (A_{\rm below}+A_{\rm above})$,
where $A_{\rm below}$ and $A_{\rm above}$ denote the box areas below and above the line of the power spectrum, respectively. The CC diagnostic defined in such a way shows several potential advantages over its counterparts, which may be helpful in future large surveys such as e-ROSITA. For example, the application of PEI is free of the effects of centroid shift of the X-ray halo caused by merger event, which may make the determination of centroid of the X-ray gas halo ambiguous\footnote{In such a case the calculation of PEI can be performed immediately after a rough estimate of $r_{500}$ is given by, e.g., the $r_{500}$ -temperature/luminosity scaling relations \citep{sun2009}}, and the corresponding calculation is straightforward, almost irrelevant to the complicated spectral analysis. We then classify the clusters in our sample with the new diagnostic parameter and the corresponding criteria listed in Table~\ref{tbl-2}, and show the results in Figure~\ref{fig6}. We find that our classification agrees very well with those achieved with the traditional CC diagnostics, which are also displayed in Figure~\ref{fig6}, except for the cases of A2319, A2667, and ZwCl0015. Possible reasons for the differences in classification are discussed below. 

{\it A2319 ---}
This is a nearby, massive merging system showing a huge radio halo that extends out to about 35\arcmin\ ($\sim 2$ Mpc; \citealt{storm15} and references therein). By examining the \Chandra\ image we find that the core region of the cluster is loose, flat, but well developed with distinct boundaries. In addition to this the core also shows a hierarchical structure with one flat core embedded in another. Using the PEI diagnostic, which is sensitive to such imaging substructures, we classify the cluster as a WCC system. This is supported by the results of \citet{ohara04}, who revealed a relatively weak cool gas clump slightly south of the X-ray peak on the \Chandra\ temperature map and identified it as a previously unknown cool core. On the other hand we find that, because the gas temperature is high and the core is loose, comparing with typical WCC clusters A2319 tends to possess a longer central cooling time and a lower concentration parameter. This is why it is classified as an NCC cluster with traditional CC diagnostics.   

{\it A2667 ---}
We classify this luminous cluster as an SCC system using the PEI diagnostic, and this is supported by the drastic gas temperature drop detected in the central 100 kpc in our deprojected spectral analysis (Fig.~\ref{fig7}). Previous \ROSAT\ observation indicated that the cluster shows strong evidence for a cooling core \citep{rizza98}, which is supported by the appearance of strong $H_{\alpha}$ and [O II]$\lambda$3727 lines from the central dominating galaxy \citep{rizza98}. Note that, using the central cooling time and concentration diagnostics the cluster can be classified as an SCC system marginally ($1\sigma$ confidence level), while using the cuspiness diagnostic the cluster is classified as a WCC system (close to an NCC system).

{\it ZwCl0015 ---}
This cluster is less studied in literature compared with the above two Abell systems. On its \Chandra\ image we find that, although the emission from the core region is sharply peaked, which leads to a lower central cooling time and slightly higher surface brightness concentration than typical NCC clusters, the X-ray emission from the core is apparently weaker than most WCC clusters. Moreover the gas temperature profile of this cluster shows very mild spatial variations and is actually consistent with a constant distribution (Fig.~\ref{fig7}). Due to these facts the cluster is classified as an NCC system using the PEI diagnostic, instead of a WCC classification suggested by the traditional CC diagnostics.

In order to investigate whether or not the PEI diagnostic is applicable to clusters located within $z \simeq 0.5$ or even at higher redshifts, we select three CC clusters (A1068, A1664 and RXCJ2014.8-2430), WCC (A2667, A795, and A963), and NCC (A2104, A2443, AC114) clusters respectively, 
each possessing an intermediate appearance in its own classification, as the templates, and create the corresponding simulated clusters that are located at $z=0.5$, $0.6$, and $0.7$, by cloning the three template clusters under the assumption of non-evolution. In the process of cloning we have assumed a typical 50 ks of \Chandra\ or \XMM{} exposure, and have taken into account the effects caused by the changes of angular and luminosity distances (see \citealt{santos08} for more details of this approach and \citealt{bouwens98} for the original application in the optical band). We find that, if spatial resolution and signal-to-noise ratios that can be achieved in typical \Chandra{} or \XMM{} observations are assumed, the PEI diagnostic can work well at $z=0.5$  if the flux of the simulated cluster is no less than $\simeq 2.8 \times 10^{-13}$ erg $ {\rm cm^{-2}}$ ${\rm s^{-1}}$ for \Chandra{} or $\simeq 1.6 \times 10^{-13}$ erg $ {\rm cm^{-2}}$ ${\rm s^{-1}}$ for \XMM{}, which roughly correspond to intermediately sized clusters with a gas temperature of $\sim 3$ keV. The comparison between the PEI classifications of the simulated clusters and the PEI classifications of the origin ones was shown in Figure 8. For simulated clusters located at $z=0.6$ the PEI diagnostic fails in about $20\%$ cases due to the limited signal-to-noise ratio and spatial resolution of a typical \Chandra{}/\XMM{} observation. Based on these results, we conclude that with current \Chandra{} and \XMM{} data, the new diagnostic can be safely applied to at least intermediate redshifts as a useful complement to the traditional diagnostics, and to even higher redshifts with higher quality data provided by future missions.

It is worth noting that the power excess begins at  $k \lesssim 0.01-0.05$ $\rm kpc^{-1}$ in $31$ clusters ($76\%$ of the sample), and at $k \lesssim 0.05-0.1$ $\rm kpc^{-1}$ for seven clusters ($17 \%$). These correspond to $\gtrsim 20-100$ kpc or $\gtrsim 10-20$ kpc in real space, scales at which the fluctuations of X-ray surface brightness, gas density, and velocity were found (e.g., \citealt{walker15}, \citealt{churazov12}, \citealt{zhuravleva15}, \citealt{rebusco05}, \citealt{rebusco06}). The fluctuations, which are speculated to be caused by either AGN feedback, or merger, or both, may contribute part of the power excess but not all, since the fluctuations should be detected at relatively low levels compared with the significance of the power excess ($\simeq 1$ magnitude for SCC, $\lesssim 0.5$ magnitude for WCC at $50$ kpc, $\simeq 0.5$ magnitude for NCC at $200$ kpc). To further evaluate this, we have attempted to use a two-dimensional $\beta$ or $2\beta$ model, which is spatially smoothed, to approximated the emission distribution of the sample clusters. We find that the power spectra calculated from the modeled clusters show similar power excess as observed. Despite this, to quantitatively answer the question about how much the gas fluctuations contribute to the observed power excess still remains an interesting task in the future.

\clearpage

\section{SUMMARY}
We propose a new CC diagnostic based on the study of the two-dimensional power spectra of the \Chandra\ X-ray images of 41 galaxy clusters ($z=0.01\sim 0.54$). By calculating the power excess index we find that the CC-NCC classification based on our new diagnostic agrees very well with those obtained by the traditional CC diagnostics. The new diagnostic can be safely applied to at least intermediate redshifts as a useful complement to the traditional diagnostics.

This work was supported by the Ministry of Science and Technology of China
(grant No. 2013CB837900),
the National Science Foundation of China
(grant Nos. 11125313, 11203017, 11433002, 61271349, and 61371147),
the Chinese Academy of Sciences
(grant No. KJZD-EW-T01),
and Science and Technology Commission of Shanghai Municipality
(grant No. 11DZ2260700).

\clearpage


\begin{deluxetable}{lccccccl}
\tablecolumns{8}
\tablewidth{0pc}
\setlength{\tabcolsep}{2pt}
\tablecaption{Basic Properties of Sample Clusters. \label{tbl-1}}
\tablehead{
\colhead{Name} &  \colhead{ObsID\tablenotemark{a}} &  \colhead{R.A.}  &  \colhead{Dec.} &  \colhead{z} &  \colhead{$T_{\rm avg}\tablenotemark{b}$} &  \colhead{$M_{500}$} &  \colhead{$r_{500}$} \\
\colhead{} & \colhead{}  & \colhead{(J2000)}  & \colhead{(J2000)} &  \colhead{}  & \colhead{(keV)}   & \colhead{($10^{14}~\rm M_{\sun}$)} &  \colhead{(kpc)}}
\startdata
A0193 & 6931 & 01:25:07.3 & 	$+$08:41:36.00 & ~0.0486 & $~3.83_{-0.18}^{+0.18} ~$ & $~1.66_{-0.28}^{+0.90} ~$ & $~816_{-50}^{+127}$  \\
A0520 & 4215 & 04:54:09.7 & $+$02:55:23.41 & ~0.1990 & $~9.04_{-0.08}^{+0.08} ~$ & $~6.78_{-0.48}^{+0.74} ~$ & $~1246_{-30}^{+44}$  \\
A0697 & 4217 & 08:42:53.3 & $+$36:20:12.00 & ~0.2820 & $12.43_{-1.34}^{+1.31}$ & 	$11.34_{-3.86}^{+6.04}$ & $~1439_{-186}^{+220}$	  \\
A0795 & 11734 & 09:24:05.3 & $+$14:10:21.00 & ~0.1359 & $~5.09_{-0.27}^{+0.27} ~$ & $~2.64_{-0.52}^{+0.36} ~$ & $~928_{-66}^{+41}$  \\
A0963 & 903 & 10:17:03.4 & $+$39:02:53.66 & ~0.2060 & $~6.59_{-0.27}^{+0.28} ~$ & $~4.20_{-0.45}^{+0.66} ~$ & $~1060_{-38}^{+53}$  \\
A0970 & 12285 & 10:17:34.3 & $-$10:42:01.00 & ~0.0587 & $~4.18_{-0.39}^{+0.39} ~$ & $~5.94_{-2.07}^{+7.75} ~$ & $~1245_{-166}^{+399}$  \\
A1068 & 1652 & 10:40:44.5 & $+$39:57:11.07 & ~0.1375 & $~5.07_{-0.24}^{+0.24} ~$ & $~3.33_{-0.39}^{+0.45} ~$ & $~1003_{-50}^{+41}$ \\
A1204 & 2205 & 11:13:20.4 & $+$17:35:40.93 & ~0.1706 & $~4.56_{-0.32}^{+0.35} ~$ & $~2.09_{-0.28}^{+0.13} ~$ & $~850_{-39}^{+17}$  \\
A1651 & 4185 & 12:59:22.3 &	 $-$04:11:44.87 & ~0.0850 & $~6.50_{-0.33}^{+0.32} ~$ &	$~6.79_{-1.71}^{+2.04} ~$ &	$~1293_{-120}^{+118}$  \\
A1664 & 7901 & 13:03:42.4 &	 $-$24:14:43.66 & ~0.1283 & $~5.35_{-0.20}^{+0.26} ~$ &	$~3.79_{-0.38}^{+0.40} ~$ &	$~1050_{-38}^{+36}$  \\
A1736 & 4186 & 13:26:52.1 & $-$27:06:33.00 & ~0.0458 & $~2.60_{-0.08}^{+0.07} ~$ & $~1.98_{-0.73}^{+0.60} ~$ & $~866_{-123}^{+80}$  \\
A1991 & 3193 & 14:54:31.5 &	 $+$18:38:32.94 & ~0.0587 & $~2.64_{-0.07}^{+0.07} ~$ & $~0.90_{-0.07}^{+0.04} ~$ & $~664_{-18}^{+9}$  \\
A2034 & 12886 & 15:10:13.1 & $+$33:31:41.00 & ~0.1130 & $~8.96_{-0.32}^{+0.32} ~$ & $~8.24_{-0.35}^{+1.20} ~$ & $~1367_{-20}^{+63}$  \\
A2061 & 10449 & 15:21:15.3 & $+$30:39:17.00 & ~0.0784 & $~5.05_{-0.17}^{+0.17} ~$ & $~3.63_{-0.52}^{+1.73} ~$ & $~1051_{-51}^{+146}$  \\
A2104 & 895 & 15:40:06.8 & $-$03:17:39.00 & ~0.1533 & $~9.24_{-0.49}^{+0.49} ~$ & $~5.45_{-0.47}^{+0.68} ~$ & $~1176_{-35}^{+47}$  \\
A2163 & 1653 & 16:15:34.1 & $-$06:07:26.00 & ~0.2030 & $16.09_{-0.53}^{+0.52}$ & $19.49_{-2.51}^{+2.07} $ & $~1770_{-79}^{+61}$  \\
A2255 & 894 & 17:12:31.0 & 	$+$64:05:33.00 & ~0.0806 & $~6.64_{-0.14}^{+0.14} ~$ & $~4.23_{-0.48}^{+0.81} ~$ & $~1105_{-44}^{+66}$  \\
A2319 & 3231 & 19:20:45.3 & $+$43:57:43.00 & ~0.0557 & $10.17_{-0.32}^{+0.32}$ & $11.50_{-4.75}^{+3.07}$ & $~1554_{-253}^{+127}$ \\
A2443 & 12257 & 22:25:07.4 & $+$17:20:17.00 & ~0.1080 & $~5.93_{-0.49}^{+0.49} ~$ & $~4.25_{-1.01}^{+2.43} ~$ & $~1097_{-94}^{+178}$  \\
A2554 & 1696 & 23:12:15.1 & $-$21:33:56.00 & ~0.1108 & $~4.54_{-0.34}^{+0.41} ~$ & $~1.83_{-0.32}^{+0.30} ~$ & $~828_{-52}^{+43}$ \\
A2657 & 4941 & 23:44:56.3 & $+$09:11:24.00 & ~0.0402 & $~3.99_{-0.12}^{+0.12} ~$ & $~2.41_{-0.90}^{+1.59} ~$ & $~927_{-135}^{+171}$  \\
A2667 & 2214 & 23:51:39.3 & $-$26:05:03.22 & ~0.2300 & $~7.97_{-0.74}^{+0.89} ~$ & $~6.51_{-1.75}^{+2.52} ~$ & $~1217_{-120}^{	+140}$  \\
A3158 & 3712 & 03:42:39.6 & $-$53:37:50.00 & ~0.0597 & $~4.96_{-0.09}^{+0.09} ~$ & $~3.22_{-0.37}^{+0.50} ~$ & $~1016_{-41}^{+49}$  \\
A3364 & 9419 & 05:47:34.2 & $-$31:53:01.00 & ~0.1483 & $~7.41_{-0.57}^{+0.57} ~$ & $~4.83_{-1.13}^{+1.19} ~$ & $~1131_{-96}^{+86}$  \\
A3376 & 3202 & 06:02:10.1 & $-$39:57:22.31 &	~0.0456 & $~4.37_{-0.11}^{+0.14} ~$ & $~3.35_{-0.58}^{+0.85} ~$ &	$~1033_{	-62}^{+81}$  \\
A3391 & 4943 & 06:26:15.4 & $-$53:40:52.00 & ~0.0514 & $~4.88_{-0.21}^{-0.20} ~$ & $~2.27_{-0.43}^{+0.74} ~$ & $~905_{-62}^{+89}$  \\
A3395SW & 4944 & 06:26:48.0 & $-$54:32:43.00 & ~0.0510 & $~4.71_{-0.23}^{+0.24} ~$ & $~2.10_{-0.27}^{+0.56} ~$ & $~883_{-39}^{+73}$  \\
A3822 & 8269 & 21:54:06.2 & $-$57:50:49.00 & ~0.0759 & $~5.20_{-0.25}^{+0.26} ~$ & $~3.37_{-0.49}^{+0.62} ~$ & $~1026_{-52}^{+59}$  \\
AC114 & 1562 & 22:58:52.3 & $-$34:46:55.00 & ~0.3120 & $~7.38_{-0.38}^{+0.38} ~$ & $~3.33_{-0.18}^{+0.17} ~$ & $~947_{-17}^{+15}$  \\
ESO306-G170B & 3188 & 05:40:06.4 & $-$40:50:08.73 & ~0.0358 & $~2.66_{-0.11}^{+0.13} ~$ & $~0.87_{-0.14}^{+0.21} ~$ & $~662_{-36}^{+50}$  \\
IC1262	 & 7322 & 17:33:02.8 & $+$43:45:44.07 & ~0.0344 & $~2.28_{-0.07}^{+0.07} ~$ & $~0.43_{-0.03}^{+0.05} ~$ & $~522_{-12}^{+18}$  \\
MACSJ2211.7-0349 & 3284 & 22:11:44.6 & $-$03:49:47.00 & ~0.2700 & $14.93_{-2.32}^{+2.81}$ & $~9.32_{-0.92}^{+1.19} ~$ & $~1354_{-46}^{+56}$  \\
NGC1550 & 3186 & 04:19:37.9 & $+$02:24:31.95 & ~0.0120 & $~1.27_{-0.03}^{+0.02} ~$ & $~0.26_{-0.03}^{+0.10} ~$ & $~446_{-15}^{+51}$  \\
PKS0745-19 & 6103 & 07:47:31.4 & $-$19:17:42.29 & ~0.1028 & $~8.93_{-0.51}^{+0.66} ~$ & $~7.11_{-0.56}^{+0.72} ~$ & $~1305_{-35}^{+43}$  \\
RBS797	 & 7902 & 09:47:12.8 & $+$76:23:13.77 & ~0.3540 & $~9.69_{-0.83}^{+0.82} ~$ & $~6.45_{-1.29}^{+1.50} ~$ & $~1164_{-84}^{+84}$  \\
RXCJ1524-3154 & 9401 & 15:24:12.9 & $-$31:54:21.99 & ~0.1028 & $~4.22_{-0.16}^{+0.22} ~$ & $~3.13_{-0.38}^{+0.35} ~$ & $~993_{-42}^{+35}$  \\
RXCJ2014.8-2430 & 11757 & 20:14:51.6 & $-$24:30:22.52 & ~0.1612 & $~7.14_{-0.40}^{+0.44} ~$ & $~4.15_{-0.40}^{+0.51} ~$ & $~1072_{-35}^{+42}$  \\
RXJ1423.8+2404 & 4195 & 14:23:47.9 & $+$24:04:42.37 & ~0.5431 & $~7.67_{-0.51}^{+0.51} ~$ & $~3.76_{-0.80}^{+0.69} ~$ & $~907_{-69}^{+52}$  \\
Zw3146 & 909 & 10:23:39.6 & $+$04:11:11.90 & ~0.2906 & $~9.26_{-0.61}^{+0.62} ~$ & $~6.64_{-0.55}^{+0.97} ~$ & $~1200_{-34}^{+56}$  \\
ZwCl0015 & 12251 & 00:06:20.6 & $+$10:51:52.98 & ~0.1675 & $~6.85_{-0.36}^{+0.53} ~$ & $~2.83_{-0.42}^{+0.28} ~$ & $~941_{-48}^{+30}$  \\
ZwCl2089 & 10463 & 09:00:36.9 & $+$20:53:40.27 & ~0.2400 & $~4.60_{-0.32}^{+0.34} ~$ & $~2.50_{-0.34}^{+0.35} ~$ & $~882_{-42}^{+39}$  \\
\enddata
\tablenotetext{a}{Only one time observation is used for each cluster, which has the best signal to noise ratio and the minimum offset.}
\tablenotetext{b}{Average temperature is calculated for $0.2-0.5$ $r_{500}$.}
\end{deluxetable}


\begin{table}
\begin{center}
\centering
\caption{Three traditional cool core diagnostics and the new one (i.e., PEI) introduced in this work. \label{tbl-2}}
\setlength{\tabcolsep}{15pt}
\renewcommand{\arraystretch}{1}
\begin{tabular}{lcccc}
\tableline\tableline
Category &   CCT    &   Cuspiness   &   Concentration  & PEI \\
         &   $t_{\rm cool}$ $(h^{-1/2}_{71}~\rm Gyr)$  &   $\alpha$    &   $C_{\rm SB}$  & $\Delta_{\rm PE}$  \\
\tableline
$\rm SCC$  &  $< 1$  &  $>0.75$  &  $>0.155$  &  $>0.42$\\
\tableline
$\rm WCC$  &  $1 - 7.7$  &  $0.5-0.75$  &  $0.075-0.155$  &  $0.31-0.42$\\
\tableline
$\rm NCC$  &  $> 7.7$  &  $<0.5$  &  $<0.075$  &  $<0.31$\\
\tableline
\end{tabular}
\end{center}
\end{table}


\begin{deluxetable}{llcccccc}
\tablecolumns{7}
\tablewidth{0pc}
\setlength{\tabcolsep}{2pt}
\tabletypesize{\footnotesize}
\tablecaption{Classifications of sample clusters with both the three traditional diagnostics (coincidentally) and the new CC diagnostic.
\label{tbl-3}}
\tablehead{
\colhead{Name\tablenotemark{1}} &  \colhead{$t_{\rm cool}$\tablenotemark{2}}&  \colhead{$\alpha$\tablenotemark{3}}  &  \colhead{$C_{\rm SB}$\tablenotemark{4}} &  \colhead{$r_{\rm cool}$\tablenotemark{5}} &  \colhead{$\Delta_{\rm PE}$\tablenotemark{6}} & \colhead{$R_{\rm excess}$\tablenotemark{7}} & \colhead{Category\tablenotemark{8}}  \\
\colhead{} & \colhead{$(h^{-1/2}_{71}~\rm Gyr)$}  & \colhead{}  & \colhead{} &  \colhead{(kpc)}  & \colhead{} & \colhead{(\%)} & \colhead{}
 }
\startdata
A0193 & $12.22_{-2.31}^{+2.27}$ & $0.23_{-0.01}^{+0.03}$ & $0.047_{-0.002}^{+0.003}$ & $14.9_{-6.4}^{+9.1}$ & $0.24 \pm 0.03$ & $1.9 \pm 0.5$ &  NCC/NCC  \\
A0520 & $8.04_{-2.22}^{+2.21}$ & $0.03_{-0.00}^{+0.00}$ & $0.016_{-0.002}^{+0.003}$ &         ---	 & $0.27 \pm 0.04$ & --- & NCC/NCC  \\
A0697 & $11.04_{-4.62}^{+3.02}$ & $0.20_{-0.03}^{+0.02}$ & $0.035_{-0.003}^{+0.002}$ &          --- & $0.17 \pm 0.05$ & --- & NCC/NCC  \\
A0795 & $3.87_{-0.74}^{+0.42}$ & $0.68_{-0.04}^{+0.02}$ & $0.120_{-0.005}^{+0.006}$ & $83.6_{-7.2}^{+13.5}$ & $0.38 \pm 0.02$ & --- & WCC/WCC  \\
A0963 & $2.32_{-0.29}^{+0.27}$ & $0.52_{-0.03}^{+0.03}$ & $0.098_{-0.005}^{+0.005}$ & $88.1_{-9.9}^{+12.2}$ & $0.36 \pm 0.02$ & $1.7 \pm 0.2$ & WCC/WCC  \\
A0970 & $15.45_{-4.44}^{+3.47}$ & $0.20_{-0.02}^{+0.03}$ & $0.041_{-0.003}^{+0.003}$ &          --- & $0.26 \pm 0.06$ & --- & NCC/NCC  \\
A1068~~ & $0.91_{-0.07}^{+0.06}$ & $1.09_{-0.04}^{+0.03}$ & $0.281_{-0.010}^{+0.011}$ & $109.4_{-5.3}^{+4.4}$ & $0.56 \pm 0.01$ & $45.7 \pm 21.9$ &  SCC/SCC \\
A1204 & $0.75_{-0.08}^{+0.07}$ & $1.12_{-0.04}^{+0.02}$ & $0.328_{-0.013}^{+0.013}$ & $111.6_{-5.4}^{+8.0}$ & $0.43 \pm 0.02$ & --- &  SCC/SCC \\
A1651 & $3.06_{-0.60}^{+0.60}$ & $0.70_{-0.06}^{+0.05}$ & $0.076_{-0.004}^{+0.004}$ & $66.8_{-14.7}^{+18.9}$ & $0.34 \pm 0.03$ & $10.2 \pm 1.7$ &  WCC/WCC \\
A1664 & $0.99_{-0.07}^{+0.06}$ & $1.14_{-0.04}^{+0.03}$ & $0.209_{-0.008}^{+0.008}$ & $93.4_{-3.6}^{+3.8}$ & $0.52 \pm 0.01$ & $15.3 \pm 2.7$ &  SCC/SCC \\
A1736 & $24.62_{-10.89}^{+6.50}$ & $0.15_{-0.03}^{+0.01}$ & $0.022_{-0.002}^{+0.002}$ &         --- & $0.22 \pm 0.08$ & $1.2 \pm 0.3$ &  NCC/NCC  \\
A1991 & $0.67_{-0.02}^{+0.01}$ & $1.16_{-0.03}^{+0.01}$ & $0.204_{-0.007}^{+0.007}$ & $67.3_{-1.3}^{+1.4}$ & $0.56 \pm 0.01$ & $50.4 \pm 11.7$ & SCC/SCC \\
A2034 & $19.79_{-3.09}^{+2.97}$ & $0.12_{-0.00}^{+0.01}$ & $0.030_{-0.002}^{+0.001}$ &           --- & $0.25 \pm 0.02$  & $0.3 \pm 0.0$ &  NCC/NCC  \\
A2061 & $27.75_{-11.16}^{+10.82}$ & $0.03_{-0.01}^{+0.01}$ & $0.016_{-0.002}^{+0.002}$ &    --- & $0.16 \pm 0.09$ & --- &  NCC/NCC  \\
A2104 & $27.77_{-5.75}^{+4.58}$ & $0.12_{-0.00}^{+0.00}$ & $0.040_{-0.002}^{+0.002}$ & $9.5_{-9.5}^{+8.1}$ & $0.26 \pm 0.03$ & $0.5 \pm 0.1$ & NCC/NCC  \\
A2163 & $14.36_{-2.04}^{+1.25}$ & $0.15_{-0.01}^{+0.01}$ & $0.023_{-0.001}^{+0.001}$ &      --- & $0.23 \pm 0.02$ & --- &  NCC/NCC  \\
A2255 & $28.27_{-6.21}^{+3.81}$ & $0.09_{-0.01}^{+0.01}$ & $0.019_{-0.001}^{+0.001}$ &         --- & $0.18 \pm 0.06$ & --- &  NCC/NCC  \\
A2319 & $13.40_{-4.11}^{+1.39}$ & $0.46_{-0.07}^{+0.02}$ & $0.043_{-0.002}^{+0.002}$ &         --- & $0.35 \pm 0.04$ & $0.8 \pm 0.1$ &  NCC/WCC  \\
A2443 & $14.51_{-6.38}^{+5.54}$ & $0.16_{-0.01}^{+0.02}$ & $0.043_{-0.003}^{+0.003}$ &           --- & $0.24 \pm 0.04$  & --- &  NCC/NCC  \\
A2554 & $11.87_{-3.12}^{+2.20}$ & $0.20_{-0.01}^{+0.01}$ & $0.066_{-0.004}^{+0.004}$ & 	        --- & $0.21 \pm 0.03$ & $1.1 \pm 0.3$ & NCC/NCC  \\
A2657 & $3.33_{-0.79}^{+0.84}$ & $0.61_{-0.11}^{+0.12}$ & $0.077_{-0.003}^{+0.004}$ & $38.2_{-12.6}^{+14.9}$ & $0.35 \pm 0.03$ & $1.0 \pm 0.1$ & WCC/WCC  \\
A2667 & $1.20_{-0.18}^{+0.18}$ & $0.54_{-0.04}^{+0.03}$ & $0.152_{-0.008}^{+0.007}$ & $135.9_{-12.9}^{+14.2}$ & $0.47 \pm 0.02$ & $4.5 \pm 0.5$ & WCC/SCC \\
A3158 & $11.27_{-1.54}^{+1.28}$ & $0.29_{-0.01}^{+0.02}$ & $0.041_{-0.002}^{+0.002}$ & $24.8_{-3.9}^{+7.0}$ & $0.25 \pm 0.02$ & $0.1 \pm 0.0$ & NCC/NCC  \\
A3364 & $13.19_{-5.65}^{+3.77}$ & $0.14_{-0.01}^{+0.01}$ & $0.040_{-0.004}^{+0.003}$ & $13.5_{-13.5}^{+46.4}$ & $0.17 \pm 0.04$ & --- &  NCC/NCC  \\
A3376 & $9.44_{-0.94}^{+1.10}$ & $0.27_{-0.01}^{+0.01}$ & $0.027_{-0.002}^{+0.002}$ &
--- & $0.29 \pm 0.03$ & --- & NCC/NCC \\
A3391 & $24.47_{-7.37}^{+6.87}$ & $0.14_{-0.01}^{+0.01}$ & $0.037_{-0.002}^{+0.002}$ &  	    --- & $0.19 \pm 0.05$ & $0.7 \pm 0.2$ & NCC/NCC  \\
A3395SW & $20.14_{-3.74}^{+2.56}$ & $0.31_{-0.02}^{+0.01}$ & $0.039_{-0.003}^{+0.002}$ & 	    --- & $0.18 \pm 0.04$ & --- & NCC/NCC  \\
A3822 & $9.56_{-2.51}^{+2.21}$ & $0.39_{-0.03}^{+0.04}$ & $0.037_{-0.002}^{+0.003}$ & $38.9_{-9.5}^{+15.1}$ & $0.14 \pm 0.05$ & --- & NCC/NCC  \\
AC114 & $10.71_{-1.68}^{+1.39}$ & $0.15_{-0.01}^{+0.00}$ & $0.034_{-0.002}^{+0.002}$ &  	       --- & $0.18 \pm 0.04$ & --- & NCC/NCC  \\
ESO306-G170B & $2.02_{-0.26}^{+0.34}$ & $0.52_{-0.03}^{+0.05}$ & $0.125_{-0.006}^{+0.006}$ & $45.9_{-7.5}^{+6.9}$ & $0.40 \pm 0.02$ & $20.7 \pm 2.7$ & WCC/WCC  \\
IC1262	 & $1.15_{-0.06}^{+0.08}$ & $0.69_{-0.02}^{+0.03}$ & $0.127_{-0.005}^{+0.005}$ & $49.1_{-2.8}^{+2.0}$ & $0.40 \pm 0.03$ & $4.9 \pm 0.6$ & WCC/WCC  \\
MACSJ2211.7-0349~ & $6.28_{-1.71}^{+1.39}$ & $0.71_{-0.04}^{+0.04}$ & $0.129_{-0.007}^{+0.007}$ & $81.7_{-15.4}^{+24.1}$ & $0.39 \pm 0.02$ & --- & WCC/WCC  \\
NGC1550 & $0.95_{-0.04}^{+0.04}$ & $0.97_{-0.02}^{+0.03}$ & $0.232_{-0.008}^{+0.008}$ & $33.0_{-1.6}^{+1.0}$ & $0.56 \pm 0.06$ & $19.5 \pm 3.4$ & SCC/SCC  \\
PKS0745-19 & $1.00_{-0.07}^{+0.06}$ & $1.41_{-0.04}^{+0.02}$ & $0.204_{-0.007}^{+0.008}$ & $104.6_{-4.1}^{+5.3}$ & $0.56 \pm 0.01$ & $13.9 \pm 1.0$ & SCC/SCC  \\
RBS797	 & $0.86_{-0.10}^{+0.10}$ & $1.20_{-0.07}^{+0.07}$ & $0.286_{-0.011}^{+0.010}$ & $138.1_{-8.0}^{+10.0}$ & $0.49 \pm 0.01$ & --- & SCC/SCC  \\
RXCJ1524-3154 & $0.88_{-0.06}^{+0.06}$ & $1.80_{-0.06}^{+0.07}$ & $0.321_{-0.011}^{+0.011}$ & $79.1_{-3.1}^{+3.2}$ & $0.61 \pm 0.01$ & --- & SCC/SCC  \\
RXCJ2014.8-2430 & $0.74_{-0.05}^{+0.06}$ & $1.76_{-0.05}^{+0.07}$ & $0.296_{-0.011}^{+0.011}$ & $104.5_{-6.1}^{+5.3}$ & $0.63 \pm 0.01$ & $37.5 \pm 2.0$ & SCC/SCC  \\
RXJ1423.8+2404 & $0.87_{-0.11}^{+0.08}$ & $1.71_{-0.12}^{+0.07}$ & $0.298_{-0.013}^{+0.012}$ & $114.6_{-6.6}^{+10.7}$ & $0.44 \pm 0.02$ & $26.1 \pm 3.6$ & SCC/SCC  \\
Zw3146 & $0.94_{-0.06}^{+0.07}$ & $0.98_{-0.02}^{+0.02}$ & $0.207_{-0.008}^{+0.008}$ & $143.0_{-6.9}^{+5.8}$ & $0.55 \pm 0.01$ & $7.6 \pm 0.5$ & SCC/SCC  \\
ZwCl0015 & $2.91_{-0.33}^{+0.32}$ & $0.61_{-0.03}^{+0.03}$ & $0.082_{-0.005}^{+0.005}$ & $49.1_{-6.0}^{+7.3}$ & $0.25 \pm 0.04$ & $4.1 \pm 0.7$ & WCC/NCC  \\
ZwCl2089~ ~& $0.79_{-0.11}^{+0.11}$ & $1.02_{-0.04}^{+0.04}$ & $0.308_{-0.012}^{+0.013}$ & $113.8_{-8.7}^{+10.7}$ & $0.53 \pm 0.02$ & $23.1 \pm 4.4$ & SCC/SCC  \\
\enddata
\tablecomments{(1) cluster name, (2) central cooling time (CCT, defined at $0.048r_{500}$), (3) cuspiness, (4) surface brightness concentration, (5) cooling radius, (6) power excess index (PEI), (7) the ratio of the central luminosity excess to the total luminosity (\S 3.2.2), (8) classification based on the three traditional diagnostics (coincidentally) and the new PEI  (see \S 3.2.4 and \S 4 for details).}
\end{deluxetable}



\begin{figure}
\centering
\epsscale{1}
\graphicspath{{figures/}}
\plotone{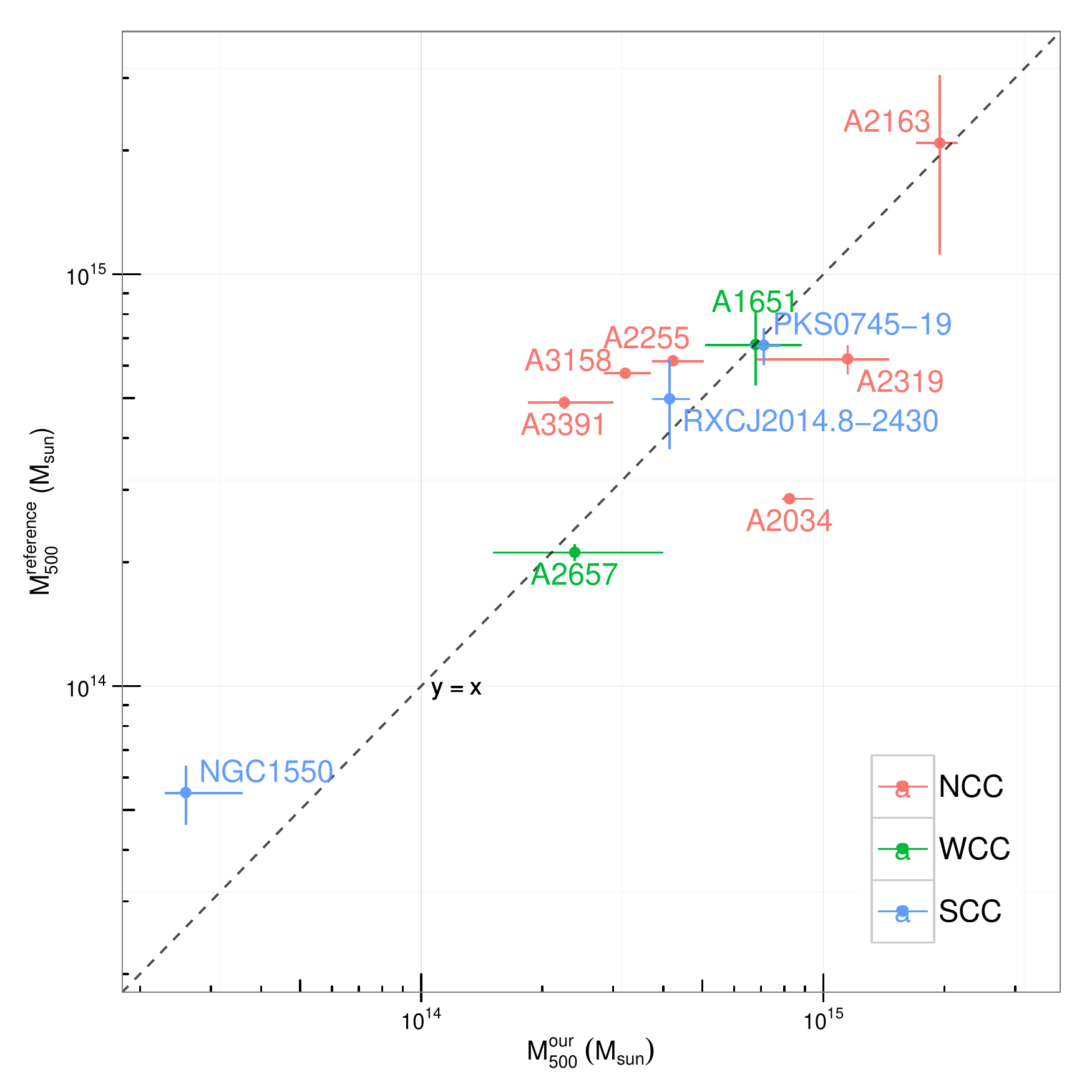}
\caption{Comparison the calculated $M_{500}$ between this work and previous work (\citealt{zhao15}). \label{fig1}}
\end{figure}

\begin{figure}
\centering
\epsscale{1}
\graphicspath{{figures/}}
\plotone{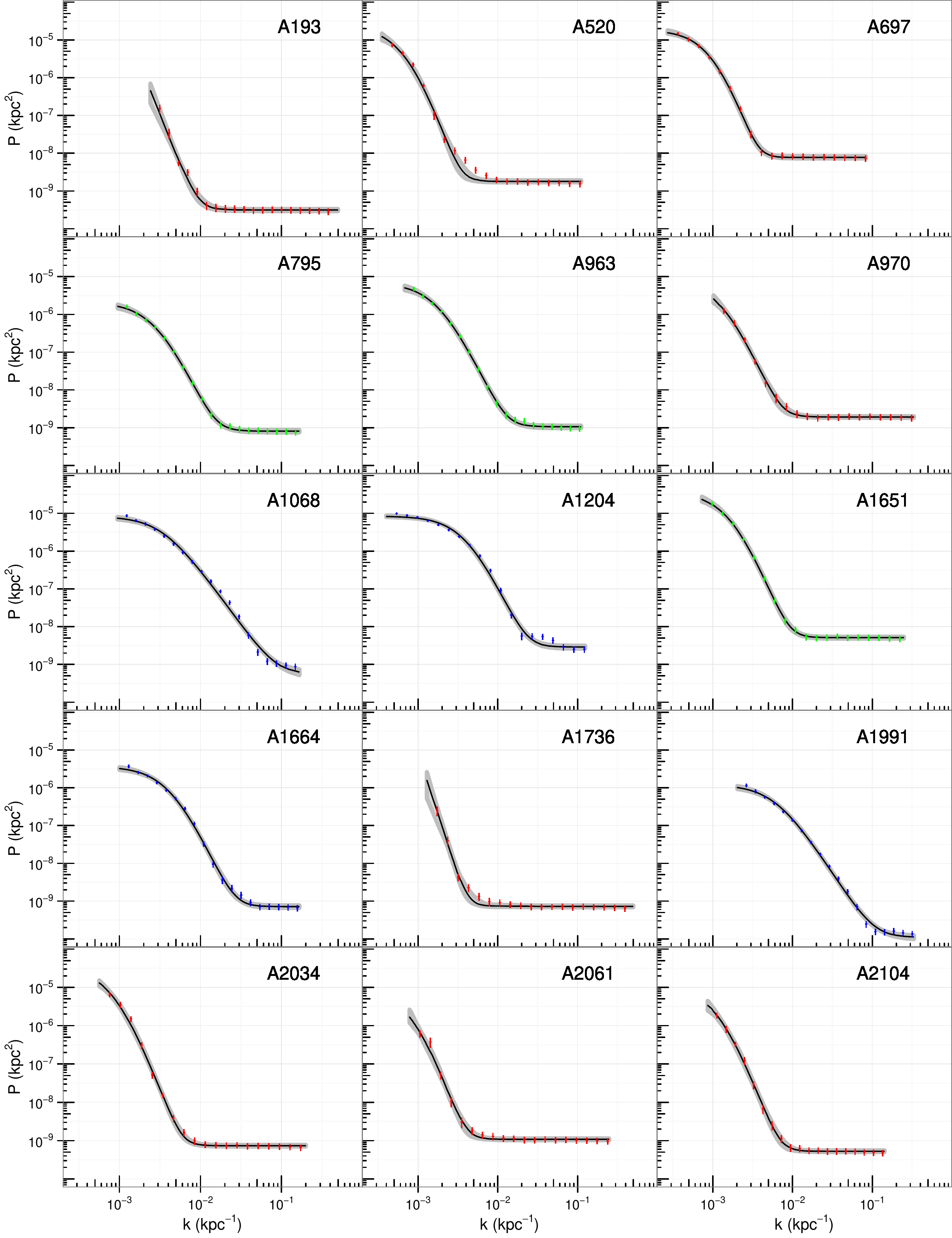}
\end{figure}
\begin{figure}
\centering
\epsscale{1}
\graphicspath{{figures/}}
\plotone{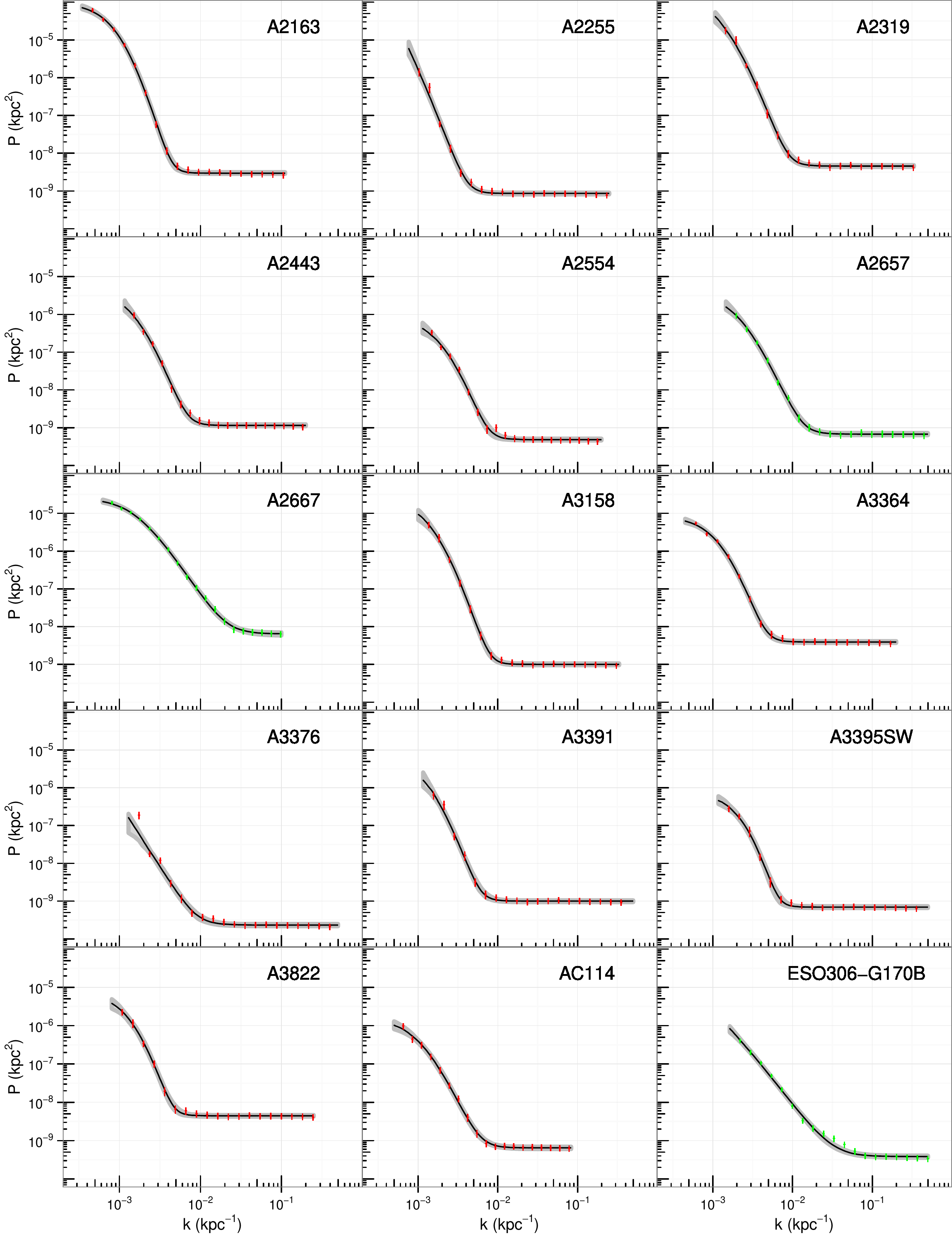}
\end{figure}
\begin{figure}
\centering
\epsscale{1}
\graphicspath{{figures/}}
\plotone{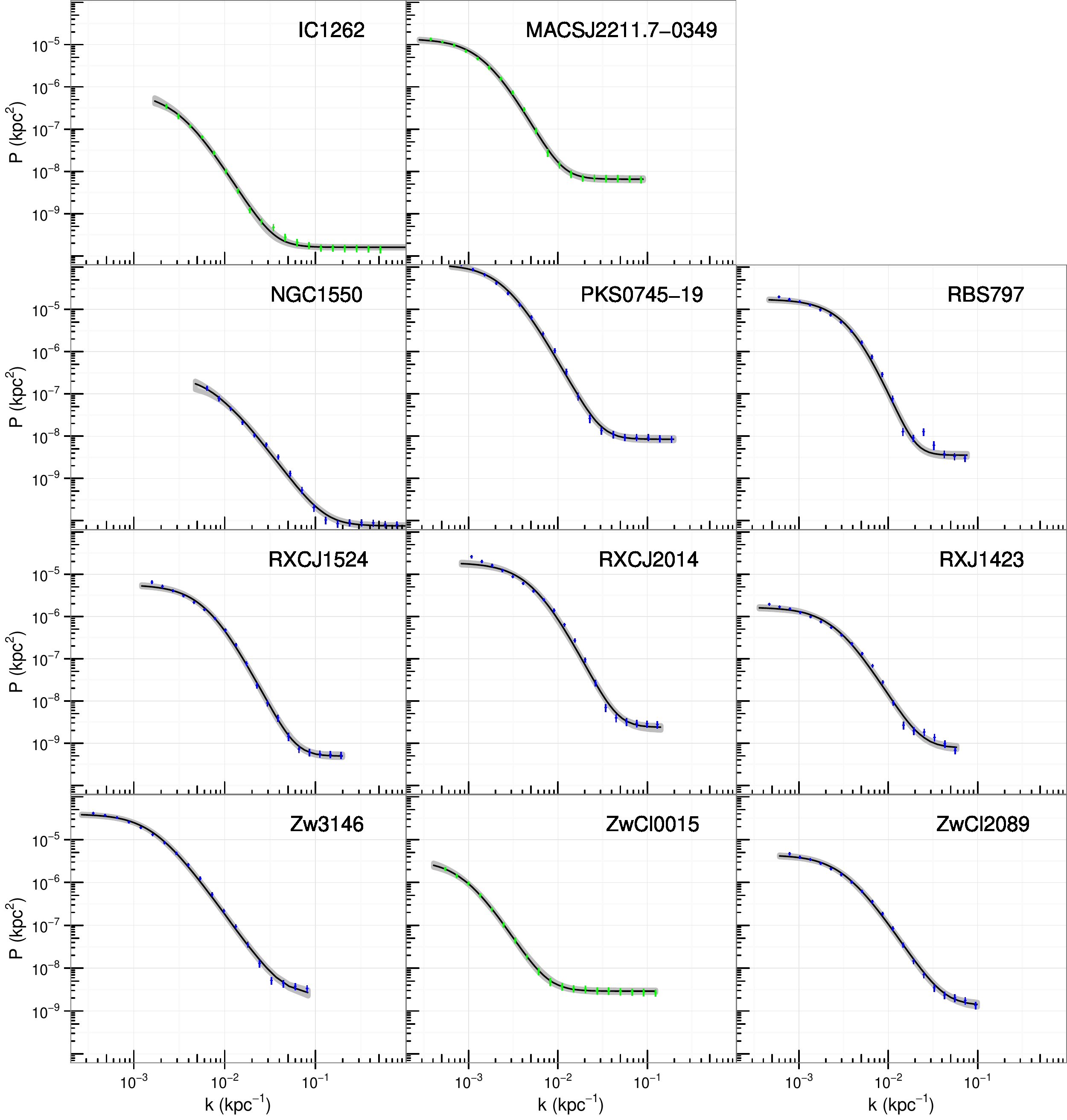}
\caption{Observed power spectra of each cluster with the best-fit model ($\beta$ + constant component) and the corresponding confidence band, which are marked in black solid line and gray shade, respectively. The cluster categories are represented with different colors, We present the SCC's in blue, WCC's in green, and NCC's in red. It should be noted that the power error bars in small wavenumber range cannot be well visualized due to the logarithmic axes, though their relative errors are typically about 10\%. \label{fig2}}
\end{figure}

\begin{figure}
\centering
\epsscale{1}
\graphicspath{{figures/}}
\plotone{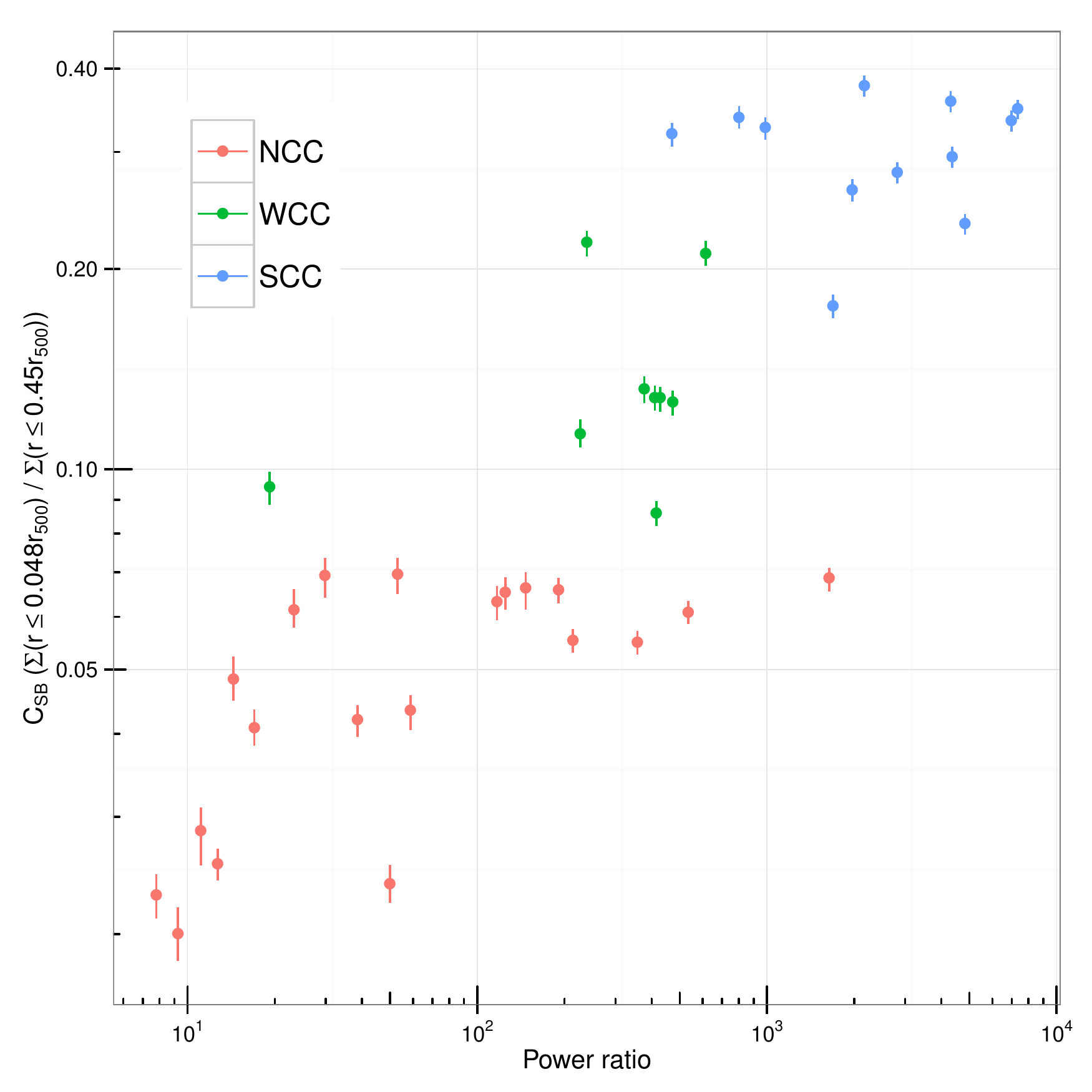}
\caption{The relative core brightness against the power ratio, see \S 3.2.5. The power ratio is defined as the ratio of power at 0.35$r_{500}$ to that at large wavenumbers where image power spectra reduce to a constant component. The SCC-WCC-NCC classifications depend on three traditional diagnostics, see \S 3.2.4\label{fig3}}
\end{figure}

\begin{figure}
\centering
\epsscale{0.8}
\graphicspath{{figures/}}
\plotone{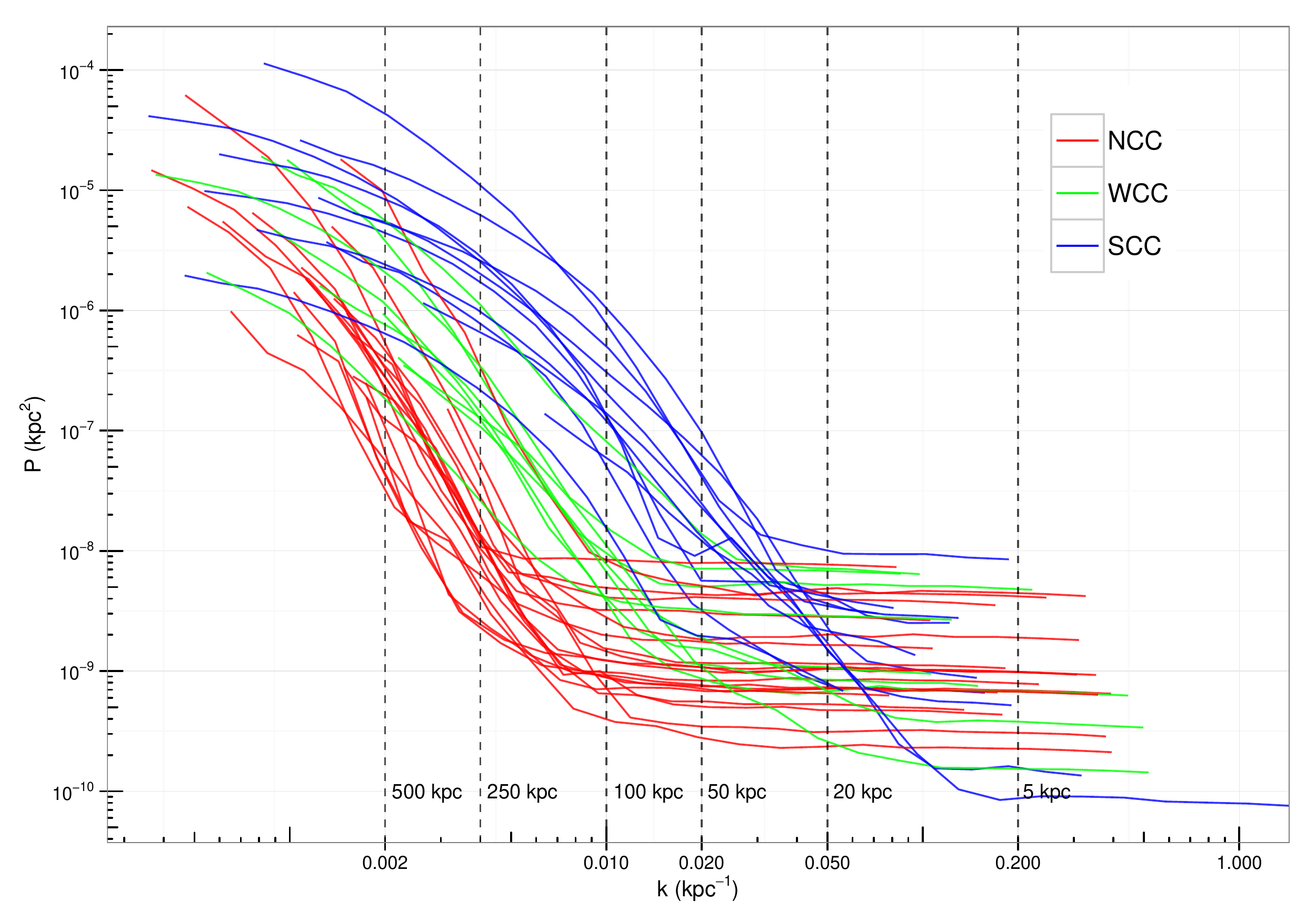}
\plotone{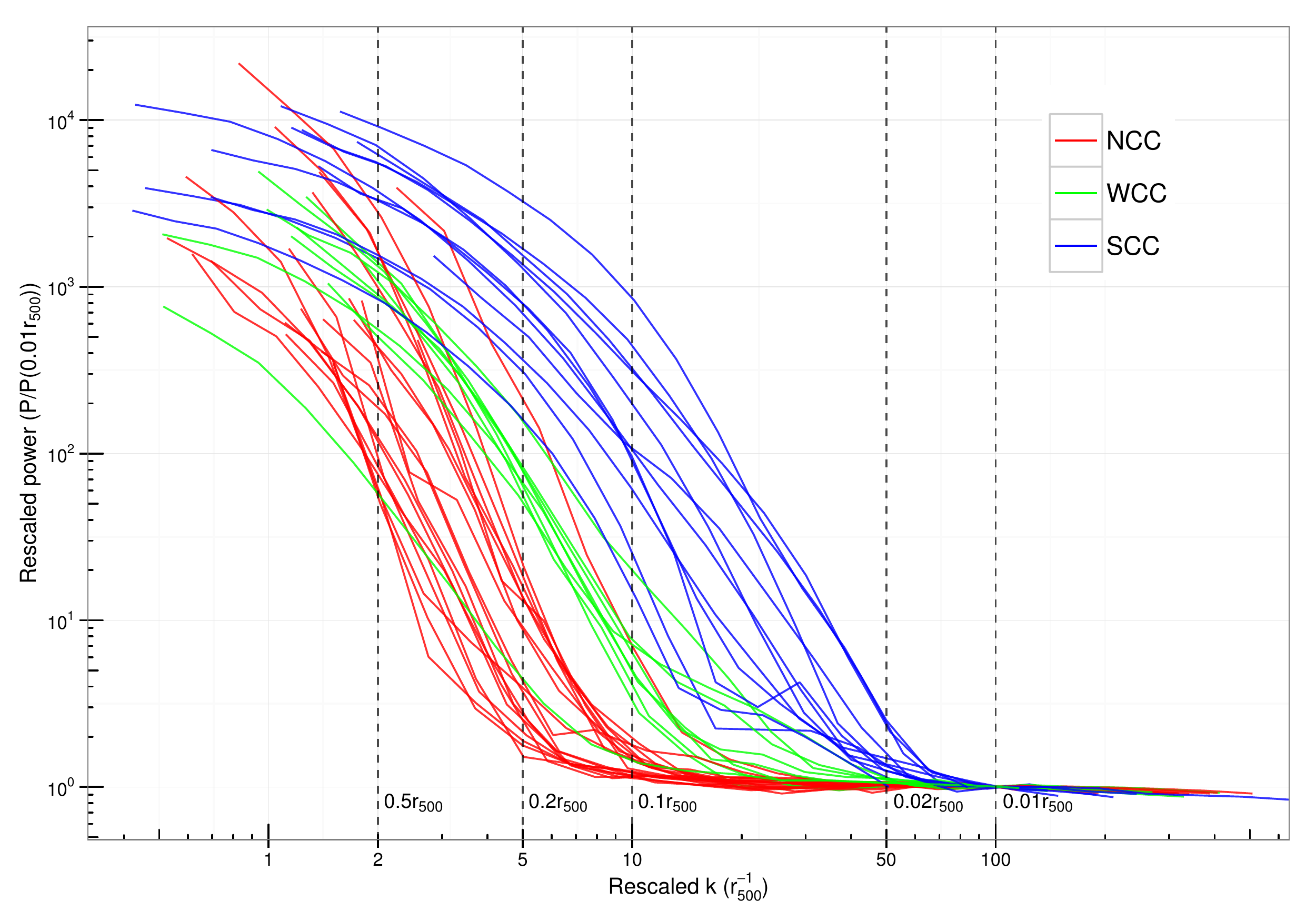}
\caption{$\it Top:$ Observed two-dimensional power spectra of all sample clusters. $\it Bottom:$ The same as top but scaled by $r_{500}$ and the power at $0.01r_{500}$, and the typical error are about 10\% for small wavenumbers and 20\% for large wavenumbers .\label{fig4}}
\end{figure}

\begin{figure}
\centering
\epsscale{1}
\graphicspath{{figures/}}
\plotone{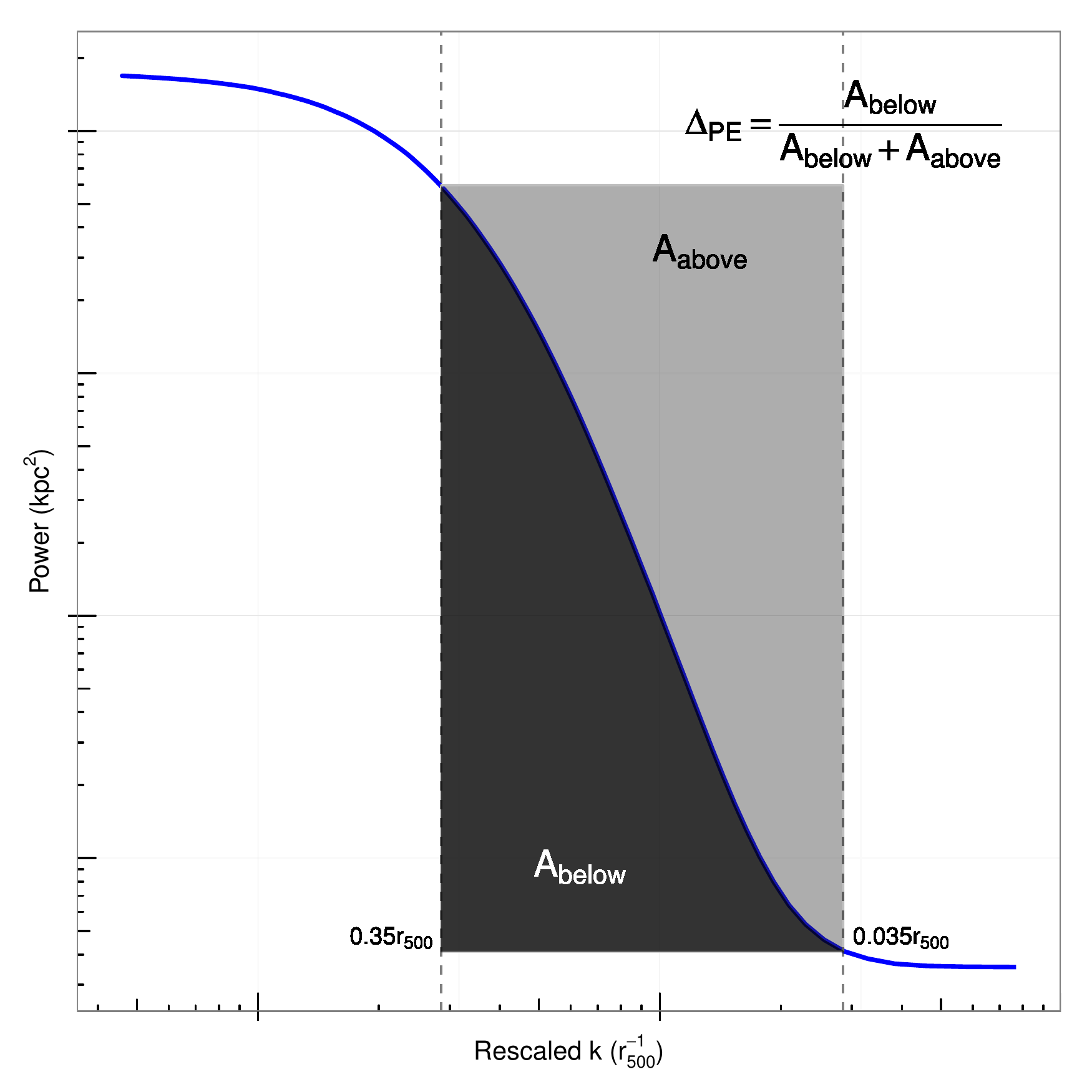}
\caption{Definition of power excess index (PEI).\label{fig5}}
\end{figure}

\begin{figure}
\centering
\epsscale{1}
\graphicspath{{figures/}}
\plottwo{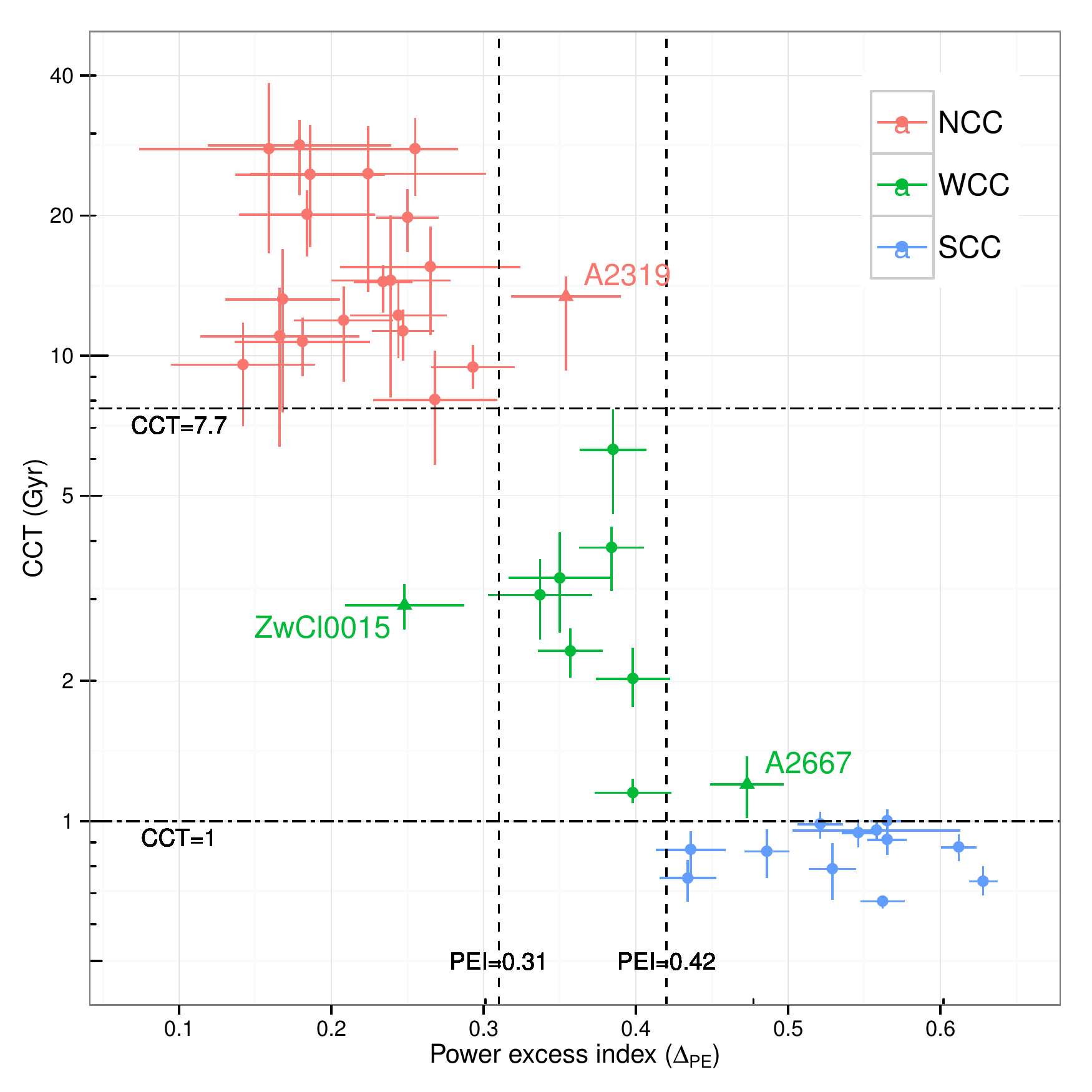}{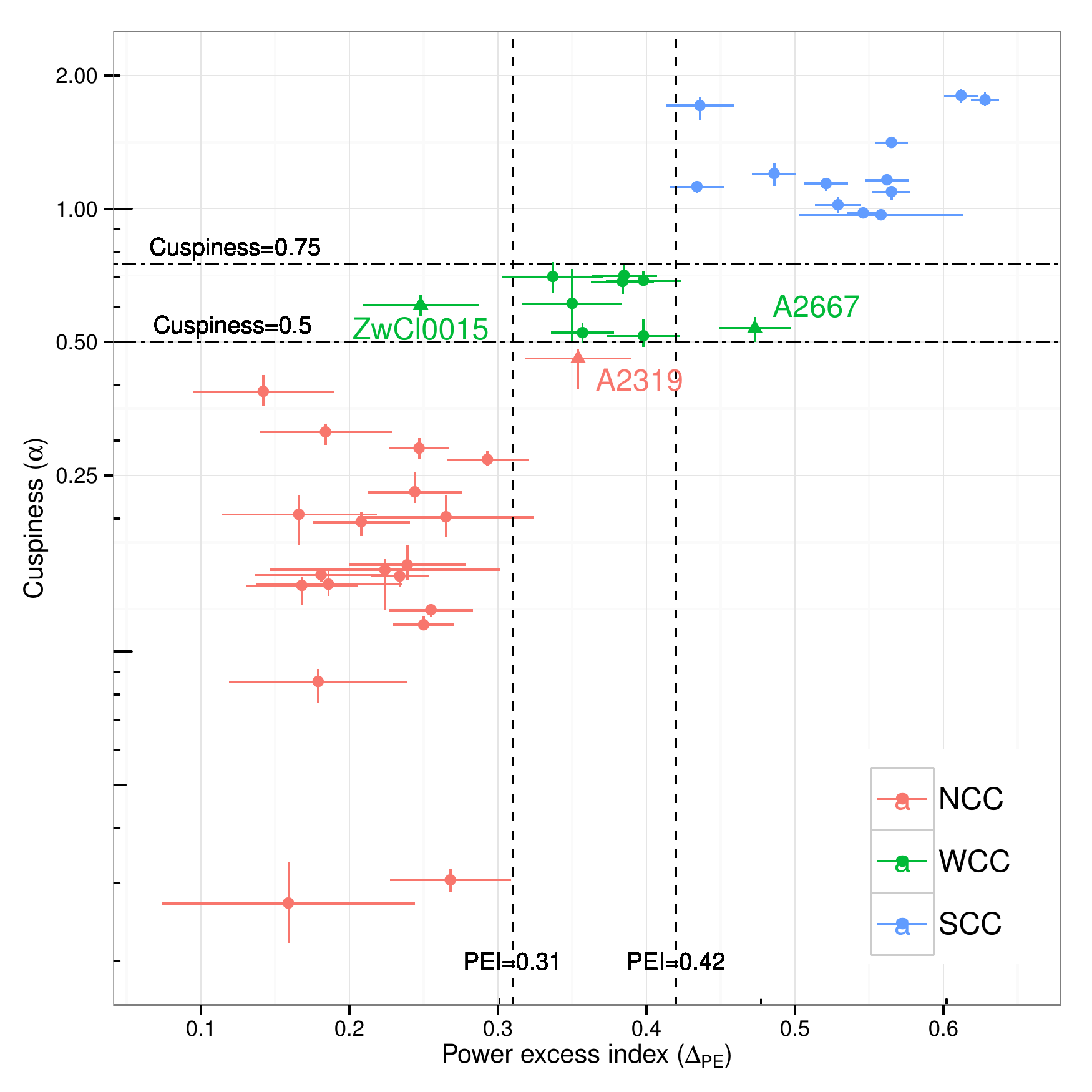}\\
\plottwo{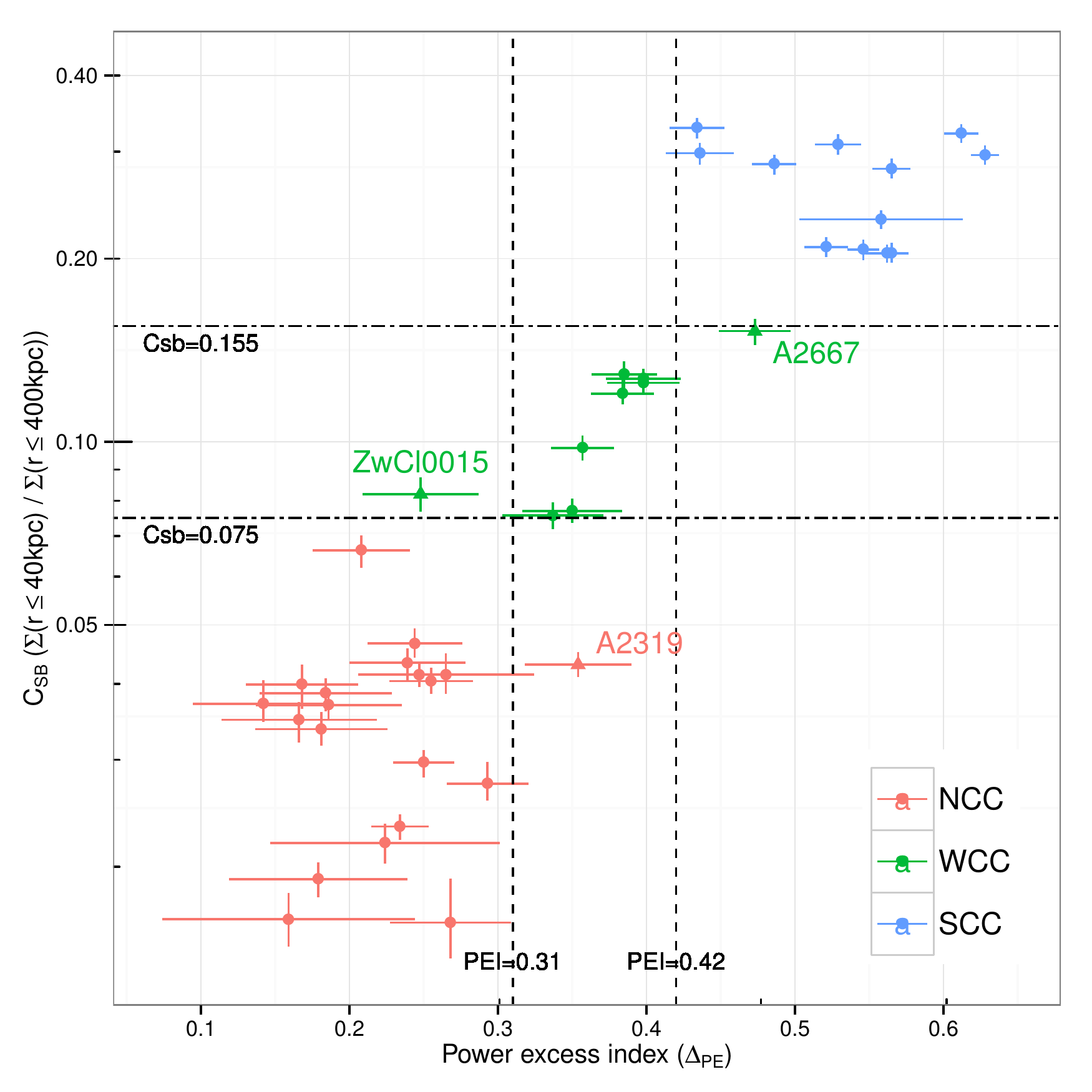}{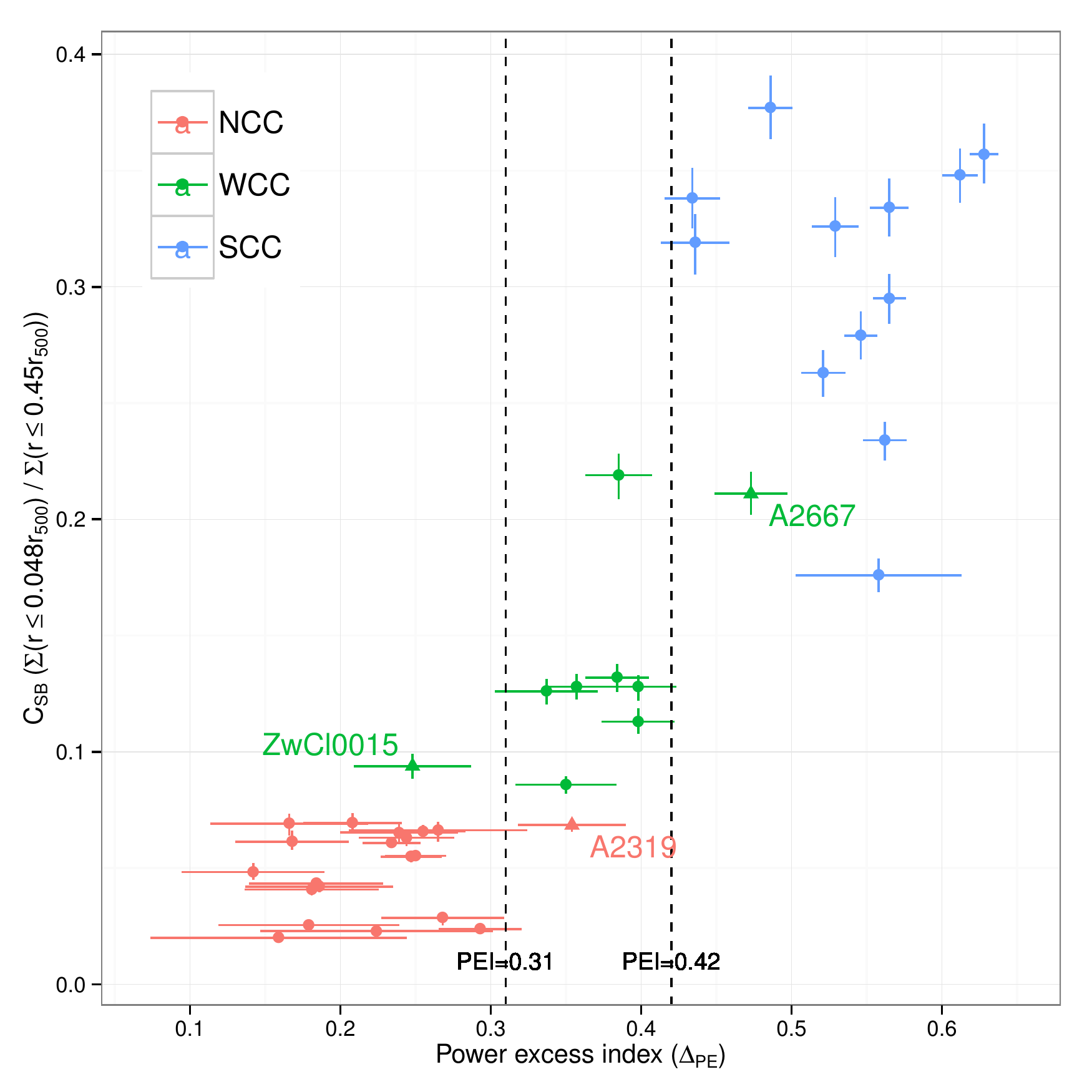}
\caption{A comparison between the SCC-WCC-NCC classifications derived with the PEI diagnostic and three traditional diagnostics. The relation between the relative core brightness (\S 3.2.5) and PEI is also plotted (lower right). \label{fig6}}
\end{figure}

\begin{figure}
\centering
\epsscale{1}
\graphicspath{{figures/}}
\plottwo{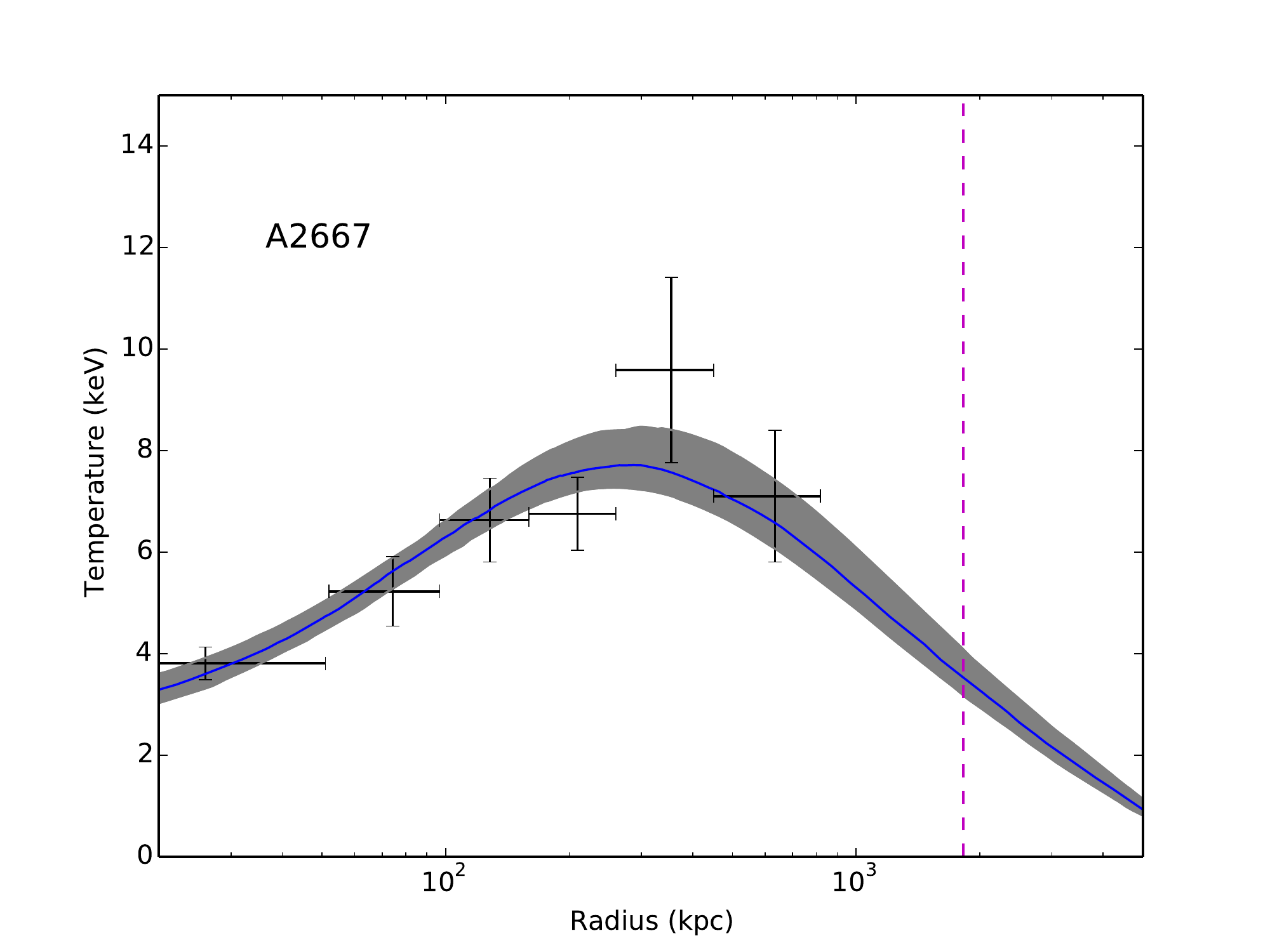}{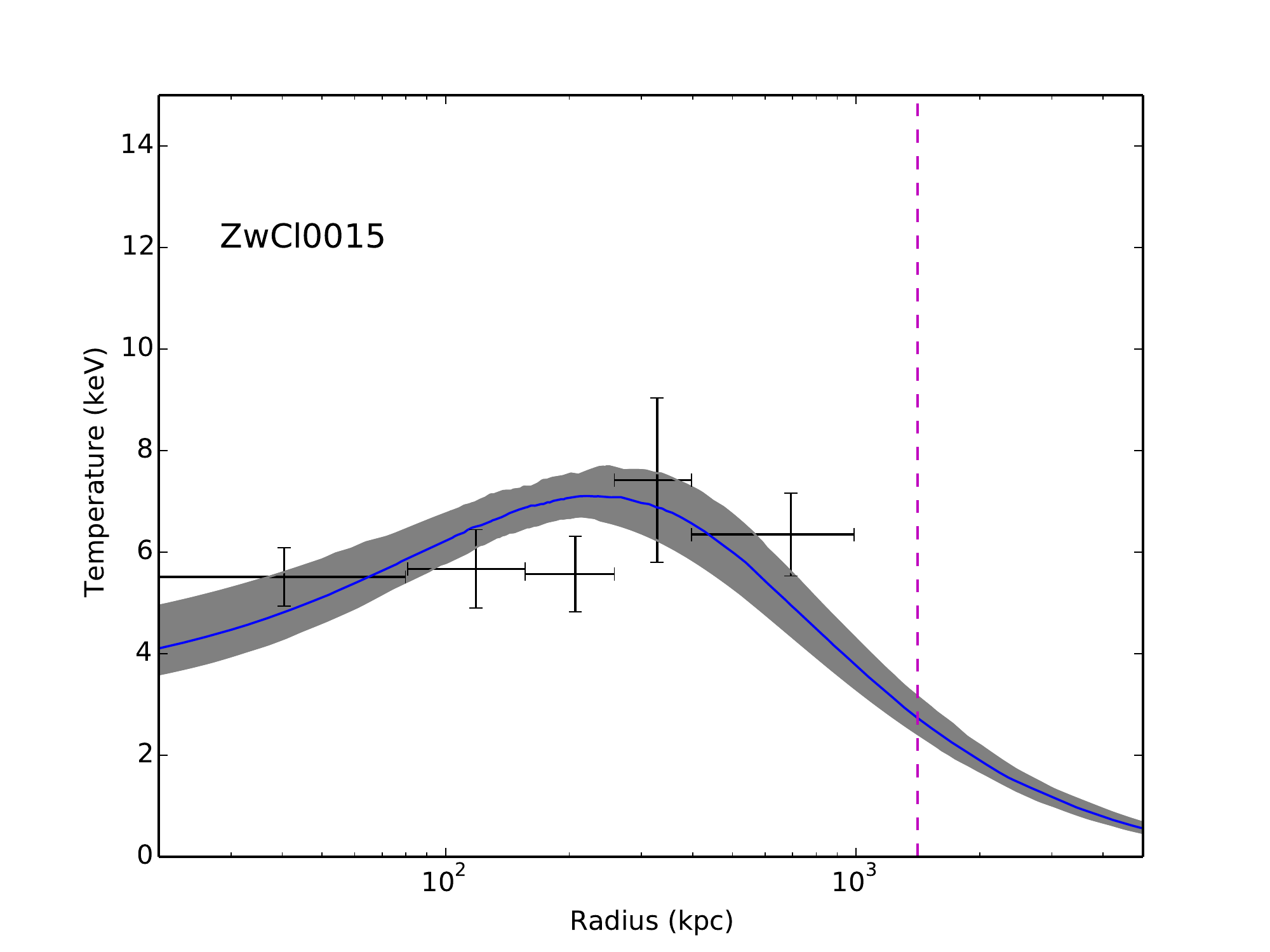}
\caption{Gas temperature profiles of A2667 and ZwCl0015 obtained with \Chandra, and the red dotted line presents the 1.5$r_{500}$.\label{fig7}}
\end{figure}

\begin{figure}
\centering
\epsscale{1}
\graphicspath{{figures/}}
\plotone{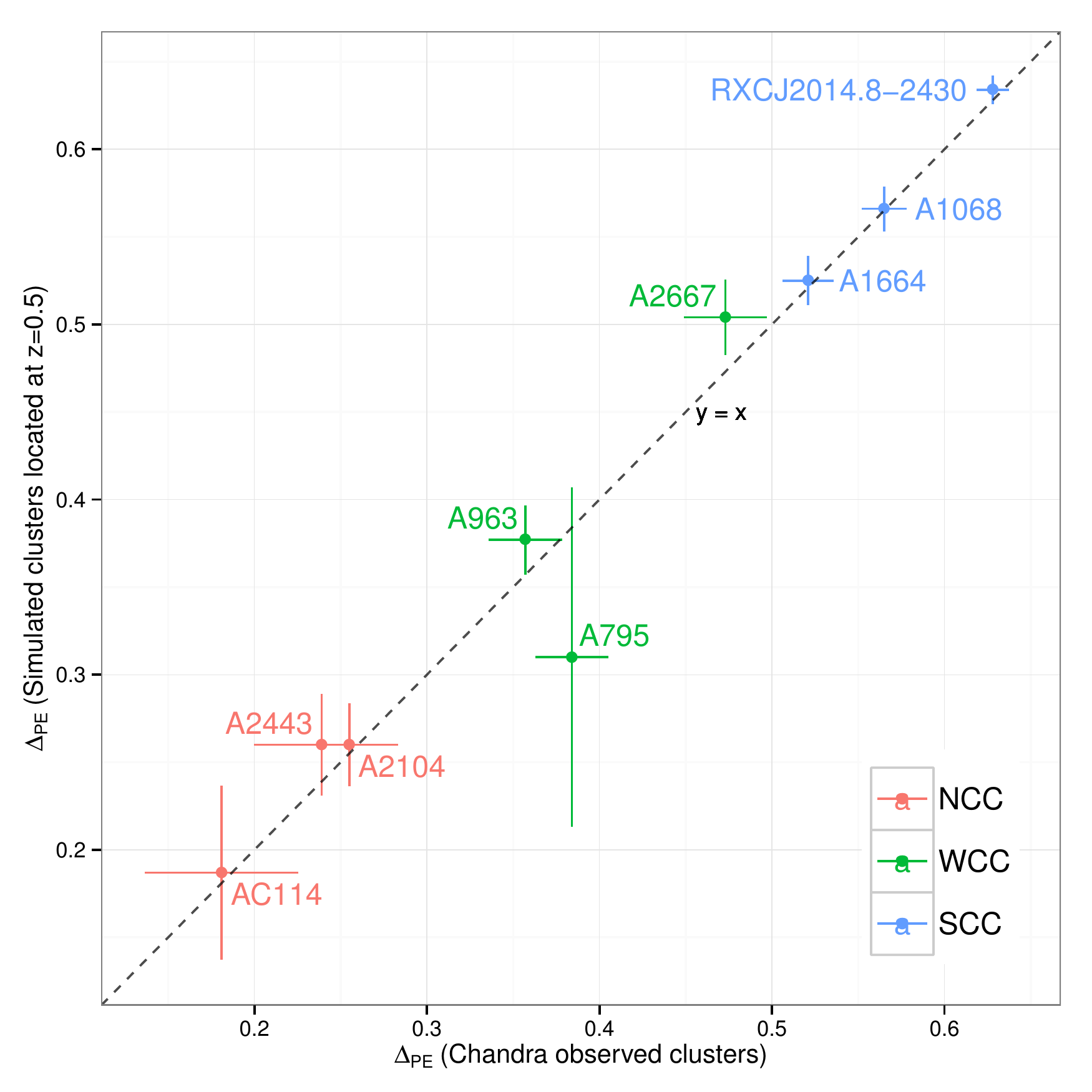}
\caption{A comparison between PEI classifications based on the observed Chandra data and the data simulated for z=0.5. \label{fig8}}
\end{figure}

\clearpage


\end{document}